\documentclass[11pt,a4paper]{article}
\usepackage{jcappub}

\usepackage{amssymb} 
\usepackage{mathrsfs} 
\usepackage{multirow} 
\usepackage{subfig} 
\usepackage{bm} 
\usepackage{graphicx}
\usepackage{amsmath}
\usepackage{amsfonts}
\usepackage[latin1]{inputenc}

\newcommand{\mincir}{\raise
  -2.truept\hbox{\rlap{\hbox{$\sim$}}\raise5.truept \hbox{$<$}\ }}
\newcommand{\magcir}{\raise
  -2.truept\hbox{\rlap{\hbox{$\sim$}}\raise5.truept \hbox{$>$}\ }}
\newcommand{\siml}{\raise
  -2.truept\hbox{\rlap{\hbox{$\sim$}}\raise5.truept \hbox{$<$}\ }}
\newcommand{\simg}{\raise
  -2.truept\hbox{\rlap{\hbox{$\sim$}}\raise5.truept \hbox{$>$}\ }}

\newcommand{\eff}{\text{eff}}

\begin{document}


\title{Constraining neutrino properties with a Euclid-like galaxy cluster survey}

\author[a,b]{M. Costanzi Alunno Cerbolini}
\author[a,b]{B. Sartoris}
\author[c,d]{Jun-Qing Xia}
\author[e]{A. Biviano}
\author[a,b,e]{S. Borgani}
\author[e,b]{M. Viel}

\affiliation[a]{Universit\'{a} di Trieste, Dipartimento di Fisica,\\ via Valerio, 2, 34127 Trieste, Italy}
\affiliation[b]{INFN-National Institute for Nuclear Physics,\\ via Valerio 2, 34127 Trieste, Italy}
\affiliation[c]{Key Laboratory of Particle Astrophysics, Institute of High Energy Physics,\\ Chinese Academy of Science, P.O.Box 918-3, Beijing 100049, P.R.China}
\affiliation[d]{Scuola Internazionale Superiore di Studi Avanzati,\\ Via Bonomea 265, I-34136 Trieste, Italy}
\affiliation[e]{INAF-Osservatorio Astronomico di Trieste,\\ via Tiepolo 11, 34143 Trieste, Italy}

\emailAdd{costanzi@oats.inaf.it}
\emailAdd{sartoris@oats.inaf.it}
\emailAdd{xiajq@ihep.ac.cn}
\emailAdd{biviano@oats.inaf.it}
\emailAdd{borgani@oats.inaf.it}
\emailAdd{viel@oats.inaf.it}

\abstract{ We perform a forecast analysis on how well a Euclid-like
  photometric galaxy cluster survey will constrain the total neutrino
  mass and effective number of neutrino species. We base our analysis
  on the Monte Carlo Markov Chains technique by combining information
  from cluster number counts and cluster power spectrum. We find that
  combining cluster data with Cosmic Microwave Background (CMB)
  measurements from Planck improves by more than an order of magnitude
  the constraint on neutrino masses compared to each probe used
  independently. For the $\Lambda$CDM+$m_\nu$ model the $2\sigma$
  upper limit on total neutrino mass shifts from $\sum m_\nu <
  0.35\text{eV}$ using cluster data alone to $\sum m_\nu <
  0.031\text{eV}$ when combined with Planck data. When a non-standard
  scenario with $N_{\eff}\neq3.046$ number of neutrino species is considered,
  we estimate an upper limit of $N_{\eff}<3.14$ ($95\%$CL), while the
  bounds on neutrino mass are relaxed to $\sum m_\nu <
  0.040\text{eV}$. This accuracy would be sufficient for a $2\sigma$
  detection of neutrino mass even in the minimal normal hierarchy
  scenario ($\sum m_\nu \simeq 0.05 \, \text{eV}$). In addition to the
  extended $\Lambda$CDM+$m_\nu$+$N_{\eff}$ model we also consider
  scenarios with a constant dark energy equation of state and a 
  non-vanishing curvature. When these models are considered the error on
  $\sum m_\nu$ is only slightly affected, while there is a larger impact 
  of the order of $\sim 15 \%$ and $\sim 20\%$ respectively on the $2\sigma$ error bar of
  $N_{\eff}$ with respect to the standard case. To assess the effect
  of an uncertain knowledge of the relation between cluster mass and
  optical richness, we also treat the $\Lambda$CDM+$m_\nu$+$N_{\eff}$
  case with free nuisance parameters, which parameterize the
  uncertainties on the cluster mass determination. Adopting the
  over-conservative assumption of no prior knowledge on the nuisance parameter
  the loss of information from cluster number counts leads to a large
  degradation of neutrino constraints. In particular, the upper bounds
  for $\sum m_\nu $ are relaxed by a factor larger than two, $\sum
  m_\nu< 0.083\,\text{eV}$ ($95\%$CL), hence compromising the
  possibility of detecting the total neutrino mass with good
  significance. We thus confirm the potential that a large
  optical/near-IR cluster survey, like that to be carried out by
  Euclid, could have in constraining neutrino properties, and we
  stress the importance of a robust measurement of masses, e.g. from
  weak lensing within the Euclid survey, in order to full exploit the
  cosmological information carried by such
  survey.\\
}


\maketitle
\section{Introduction}\label{sec_int}

Over decades neutrino oscillation experiments have provided conclusive
evidence that neutrinos have non-zero masses. Such experiments provide
constraints on the neutrino mass squared difference, while they are
not sensitive to the absolute scale of neutrino masses.  The latest
measurements, using solar, atmospheric, and reactor neutrinos, give
mass difference between neutrino species of $\Delta
m_{12}^2=7.5 \times 10^{-5}\, \text{eV}^2$ and $|\Delta
m_{23}^2|=2.3 \times 10^{-3}\, \text{eV}^2$~\citep[e.g.][]{2012PhRvD..86a3012F, 2012PhRvD..86g3012F}, which in turn
translates to a lower bound on the sum of the three masses, $\sum
m_\nu$, at $0.056 (0.095) \text{eV}$ in the normal (inverted)
hierarchy.  On the other hand, cosmological data provides a tool to
constrain neutrino masses due to the effects neutrinos induce on background
evolution and growth of structures: relativistic neutrinos
affect the Cosmic Microwave Background (CMB) anisotropies, whereas, at
low redshift when they become non-relativistic, neutrinos suppress
matter density fluctuations at small scales~\citep[e.g.][for a
review]{2002PhR...370..333D, 2006PhR...429..307L}.

Given these multiple effects massive neutrinos leave an imprint on
many cosmological observables; indeed, current neutrino constraints
from cosmology rely on a combination of data from CMB experiments,
Baryonic Acoustic Oscillations (BAOs) measurements, Supernovae
distance moduli, galaxy clustering and galaxy cluster mass function.
Recent constraints on the upper limit of the total neutrino mass lie
in the range $\sum m_\nu < 0.3 - 0.8 \,\text{eV}$ ($95\%$ 
Confidence Level (CL))~\citep[e.g.][]{2010PhRvL.105c1301T,vhs10,2012arXiv1211.3741Z, 2012JCAP...06..010X, 2012arXiv1210.2131R,
2012arXiv1212.5226H, 2012arXiv1202.0005J}, with notable exception of
Lyman$-\alpha$ data, which presents an even tighter bound of
$0.17\, \text{eV}$~\citep{2006JCAP...10..014S}.

The number of active neutrinos is known to be three to high precision
through the measurement of the invisible width of the Z boson at
LEP~\citep[$N_{\rm a} = 2.9840 \pm 0.0082$;][]{2005hep.ex....9008T},
however, the possibility remains that additional "sterile" species
exist~\citep[e.g.][]{steig13}. In fact, in order to explain the results of short baseline
neutrino oscillations experiments LSDN~\citep{2001PhRvD..64k2007A}
and \textit{MiniBooNE}~\citep{2010PhRvD..82i2005A}, as well
as the recently discovered reactor neutrino
anomaly~\citep{2011PhRvC..83e4615M, 2011PhRvD..83g3006M}, models with
one or two light sterile neutrino have been
proposed~\citep[e.g.][]{2007PhRvD..76i3005M, 2009JCAP...01..036M,
2011PhRvL.107i1801K}.  Observations of the CMB also seem to point to
the same direction, favoring the presence of extra relativistic degree
of freedom at the time of decoupling, in terms of the effective number
of neutrino species $N_{\eff}$. For instance, measurements of CMB
anisotropies from the South Pole Telescope
(SPT)~\citep{2012arXiv1212.6267H} and the Atacama Cosmology Telescope
(ACT)~\citep{2013arXiv1301.0824S}, combined with seven-year Wilkinson
Microwave Anisotropy Probe (WMAP) data~\citep{2011ApJS..192...16L} and
BAO and $H_0$ measurements, have measured $N_{\eff}=3.71 \pm 0.35$ and
$N_{\eff}=3.52 \pm 0.39$ at $68\%$ CL, respectively, thus suggesting
values higher than those expected in the canonical scenario
($N_{\eff}=3.046$).

Among the different probes of the Large Scale Structure (LSS), many works have already been
proven the ability of galaxy clusters in constraining cosmological
parameter, through both the evolution of their mass
function~\citep[e.g.][]{2001ApJ...561...13B, 2002ApJ...567..716R,
  2009ApJ...692.1060V,2010ApJ...708..645R,2010MNRAS.406.1759M,2011ARA&A..49..409A,rapetti12,benson_etal13},
and their large--scale clustering
properties~\citep[e.g.][]{borgani_guzzo01,schuecker_etal03,mana_etal13}.
Indeed, galaxy clusters supply cosmological information in different ways. The
evolution of the galaxy cluster population depends on cosmological
parameter both through the linear growth rate of density perturbations
and the redshift dependence of the volume element. Furthermore, also
clustering of galaxy clusters is sensitive to cosmological parameters
through the grow rate of perturbations. These properties makes
clusters an excellent probe of the growth of structure, and thus a
valuable tool to constrain neutrino properties~\citep[e.g.][]{2011ARA&A..49..409A, 2012ARA&A..50..353K}.

Recent constraints from galaxy cluster samples combined with CMB data,
observation of BAO and supernovae Type Ia are $\sum m_\nu \lesssim
0.52\, \text{eV}$, $N_{\eff}=
3.6_{-1.0}^{+1.4}$~\citep{2010MNRAS.406.1805M} and
$0.72 \, \text{eV}$, $N_{\eff} \lesssim 4.6$ ($95\%$
CL)~\citep{2011ARA&A..49..409A,2012AstL...38..347B}. 

While current constraints based on galaxy cluster data rely on
relatively small samples of clusters identified at redshift below one~\citep[e.g.][]{2007ApJS..172..561B,2009ApJ...699..768R},
next generation of X-ray (e.g.,
eROSITA\footnote{http://www.mpe.mpg.de/eROSITA},
WFXT\footnote{http://www.wfxt.eu/}), Sunyaev-Zeldovich (e.g.,
CCAT\footnote{http://www.ccatobservatory.org/index.cfm}, SPT-3G) and
optical (e.g., DES\footnote{http://www.darkenergysurvey.org/},
LSST\footnote{http://www.lsst.org/lsst/},
PanSTARRS\footnote{http://pan-starrs.ifa.hawaii.edu/public/}, Euclid
\footnote{http://www.euclid-ec.org/}) surveys are expected to increase
by orders of magnitude the number of galaxy clusters detected, further
extending the probed range of redshift up to $z \sim 2$. Such large
cluster surveys will provide tight constraints on cosmological
parameters, independently and complementary to those recovered from
other cosmological probes.  In this work we explore the cosmological
information contained in the cluster catalog that will be provided by
the photometric redshift survey of ESA's Euclid mission, which has
been approved for lunch in 2019.  Specifically, we will make use of
cluster number counts and cluster power spectrum to derive forecast
errors on the total neutrino mass and effective number of neutrino
species, for a Euclid-like galaxy cluster survey.  A similar analysis
based on the Fisher Matrix approach has been proposed by
\cite{2012JCAP...03..023C}. Even though the Fisher Matrix technique
has the advantage of allowing for a quick, analytic estimate of the
confidence limits, on the other hand it approximates the likelihood
function as a multivariate Gaussian function of the model
parameters. In general, this turns out to be a coarse approximation
since the likelihood function can be highly
non-Gaussian~\citep[e.g.][]{2006JCAP...10..013P}; moreover the results
obtained with this technique depend on the step chosen in the
calculation of numerical derivatives with respect to the
parameters~\citep[e.g.][]{2012JCAP...09..009W}. For these reasons we
choose to use a more robust forecast method based on the sampling of
the full likelihood function using a Monte Carlo Markov Chain (MCMC)
approach. Finally, unlike~\cite{2012JCAP...03..023C}, in our work we
explore also the non-standard scenario with $N_{\eff}>3$ and $\sum
m_\nu>0$, to assess the effects of the correlation between the two
parameters~\citep[see e.g.][]{2012arXiv1211.2154G,
  2012AstL...38..347B}.

The paper is organized as follows. In Section~\ref{sec:mnu}, we
briefly explain the physical and observable effects of $N_{\eff}$ and
$\sum m_\nu$. Section~\ref{sec:meth} describes the formalism which we
use to compute the cluster number counts (\S\ref{sub:nc}) and power
spectrum (\S\ref{sub:ps}), while in \S~\ref{sub:fore} we outline our
forecasting procedure and specify the characteristics of the galaxy
cluster survey analyzed. Our results for different cosmological model
are presented in section~\ref{sec:res}, and finally in
section~\ref{sec:conc} we draw our conclusions.

\section{Massive neutrino effects}\label{sec:mnu}
\begin{figure}
\includegraphics[width=7.3cm]{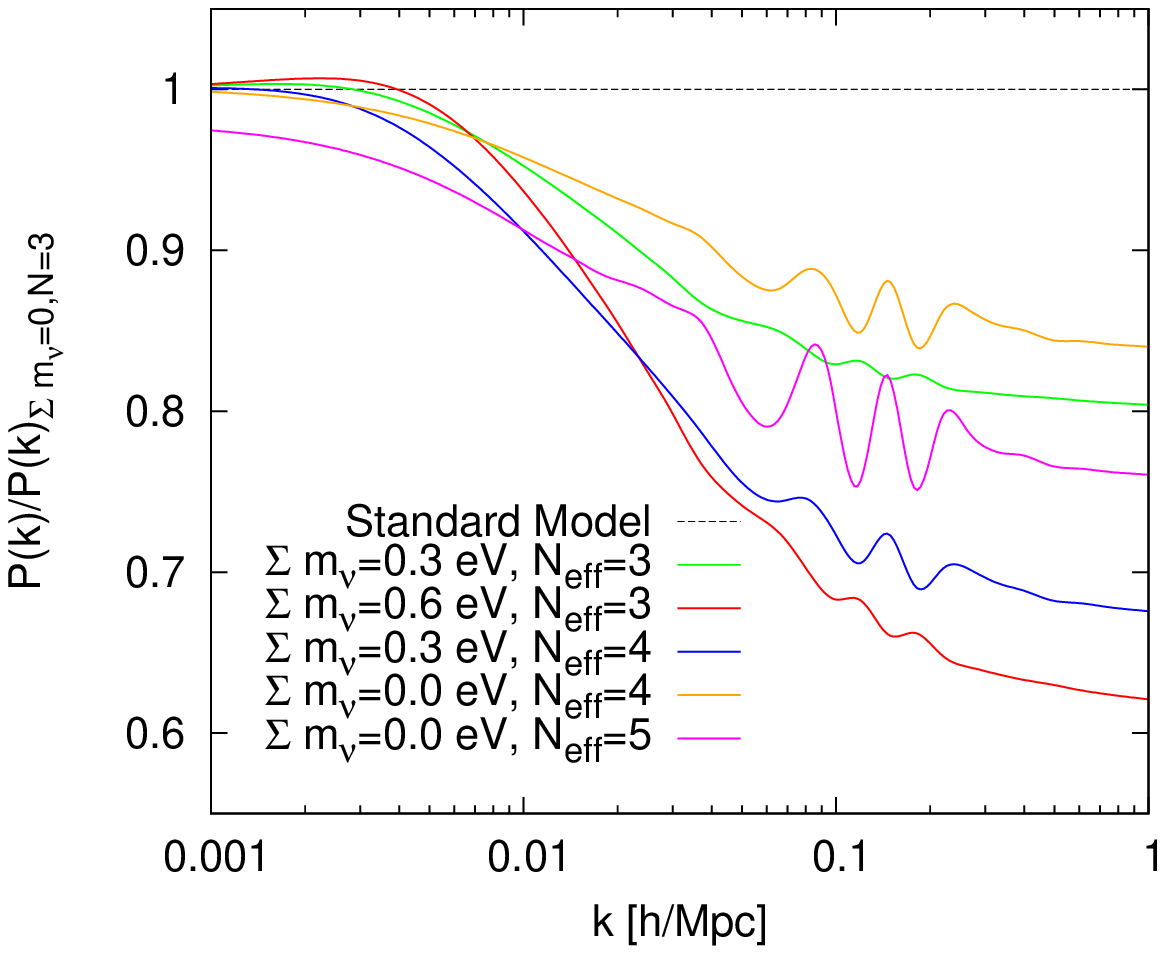} 
\includegraphics[width=7.3cm]{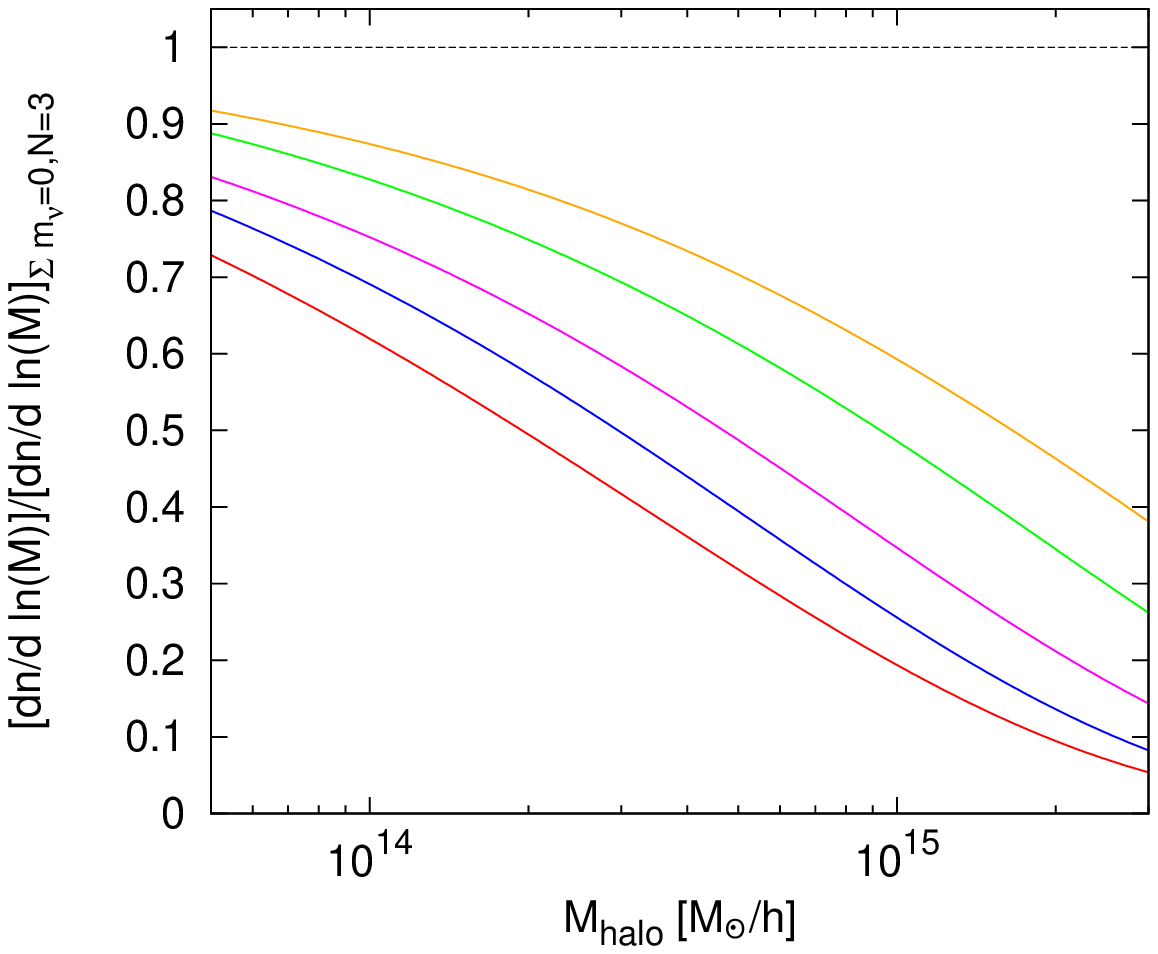}
\caption{Effects of the variation of $\sum m_\nu$ and $N_{\eff}$ on the linear matter power spectrum (left) and halo mass function (right) at $z=0$;  all other parameters $(\Omega_{\rm m}, \Omega_\Lambda, H_0, n_{\rm s}, \Delta^2_{\rm R}, \tau)$ are kept fixed to the WMAP 9-yr best-fit values for $\Lambda$CDM. See text for comments.}
\label{fig:ps_nc}
\end{figure}
In this section we briefly review the physical effects of $N_{\eff}$
and $\sum m_\nu$ and how they affect the halo mass function and the
matter power spectrum, two quantities strictly related to the
observables used to derive the forecast error (see next section).

What is actually constrained by cosmological data is not the neutrino
mass, but the neutrino energy density $\rho_\nu$, which can be related
to the neutrino mass through the relation:
\begin{equation}\label{eqn:omnu}
 \Omega_\nu= \frac{\rho_\nu}{\rho_{\rm c}}=\frac{\sum_i^{N_\nu} m_{\nu,i}} {93.14 \text{eV} h^2} \quad,
\end{equation}
where $\rho_{\rm c}$ is the critical density of the Universe, and $N_\nu$
the number of massive neutrinos. In the present work, we assume three
degenerate massive neutrino states, so that the total neutrino mass
can be written as $\sum m_\nu = 3 m_\nu$. A degenerate mass spectrum
is justified on one side by the smallness of the mass difference
measured, on the other by the incapability of the Euclid
cluster data to probe directly the neutrino mass hierarchy, as we will
explain later in \S~\ref{ssub:neff}.  The effective number of
neutrino species parameterizes the contribution of neutrinos (or any
other relativistic species) to the radiation content in the radiation
dominated era through the formula:
\begin{equation}\label{eqn:neff2}
 \rho_{\rm r} = \left[ 1 + \frac{7}{8} \left(\frac{4}{11}\right)^{4/3} N_{\eff}\right] \rho_\gamma \quad ,
\end{equation}
where $\rho_{\rm r}$ is the energy density in relativistic species and
$\rho_\gamma$ is the energy density of photons. $N_{\eff}$ can be
expressed as the sum of the number of massive neutrinos $N_\nu$ and the
contribution from extra relativistic degrees of freedom, $\Delta
N_{\eff}$. The standard value for the number of massive neutrinos is
$N_\nu=3.046$, where the $0.046$ accounts for a non-instantaneous
neutrino decoupling process and flavour neutrino oscillations
effect~\citep{2005NuPhB.729..221M}. Instead, any significant deviation
from $\Delta N_{\eff}=0$ could indicate the presence of new physics
beyond the standard model; adding an extra (thermalized) light fermion
would contribute $\Delta N_{\eff}=1$, but more generally a non-integer
$\Delta N_{\eff}$ value could arise from different physics, such as
lepton asymmetries~\citep{2012JCAP...07..025H}, partial thermalisation
of new fermions~\citep{2009JCAP...01..036M}, particle
decay~\citep{2005JHEP...09..048P}, non-thermal production of dark
matter~\citep{2012PhRvD..85f3513H}, gravity
waves~\citep{2006PhRvL..97b1301S} or early dark
energy~\citep{2011PhRvD..83l3504C}.

Modification of these parameters has effects both on the background
evolution and on structure formation.  Neutrinos decouple from the
primeval plasma at $T\approx 2 - 4 \text{MeV}$, when they are still
ultra-relativistic, and given their small mass, $\sum m_\nu <
1 \text{eV}$, they become non-relativistic (only) after
recombination. During this epoch, their energy density contributes as
radiation rather than matter, thus changing the expansion rate (via
the Friedmann equation $H^2= \frac{8 \pi G}{3}(\rho_{\rm m} + \rho_{\rm r})$) and
the time of matter-radiation equality
($a_{\text{eq}}=\rho_{\rm r}^0/(\rho_{\rm m}^0 - \rho_\nu^0)$). The epoch of
equality set the time at which sub-horizon density fluctuations start
to collapse under the action of gravity and structures can evolve.
When the other parameters are kept fixed, a larger value of $\sum
m_\nu$ or $N_{\eff}$ corresponds to a larger value of the radiation
density, and therefore a postponed time of equality.  Such
modification is seen in the matter power spectrum as a shift to larger
scale of its peak, which is determined by the size of the particle
horizon at the time of matter-radiation equality. Moreover, since on
sub-Hubble scales density fluctuations grow more efficiently during
matter dominated epoch (i.e. after equality), the matter power
spectrum is suppressed on small scales relatively to large scales 
(left panel of Fig.~\ref{fig:ps_nc}). These effects determine also a
suppression of the halo mass function (right panel of
Fig.~\ref{fig:ps_nc}).

After thermal decoupling neutrinos constitute a collisionless fluid,
whose constituent fluid elements
free-stream with a velocity, on average, equal to
their thermal velocity $v_{\rm th}$. As the Universe expands, $v_{\rm th}$
decays adiabatically till neutrinos become non-relativistic. At this
stage neutrinos behave as hot dark matter particles, suppressing
density fluctuations on scale smaller than their free-streaming
length:
\begin{equation}\label{eqn:kfs}
 k_{\text{fs}}= 0.8 \frac{\sqrt{\Omega_\Lambda + \Omega_{\rm m} (1+ z)^3}}{(1+z)^2}  \left( \frac{m}{1 \text{eV}} \right) h \text{Mpc}^{-1} \quad .
\end{equation}
   
Due to their high velocity neutrinos cannot be confined on region
smaller than their free-streaming length, thereby suppressing the
density perturbations by a factor proportional to
$(1-\Omega_\nu/\Omega_{\rm m})$. Furthermore, the absence of gravitational
back-reaction effects from free-streaming neutrinos slow down the grow
rate of CDM/baryon perturbations at late times~\citep[see
e.g.][]{2006PhR...429..307L}. Both these effects are observable in the
matter power spectrum as a suppression on small scales 
(left panel of Fig.~\ref{fig:ps_nc}), with a constant amplitude
$\Delta P(k)/P(k)\simeq -8 (\Omega_\nu/\Omega_{\rm m})$ at $k \sim 1 h
/ \text{Mpc}$, for linear structure
formation~\citep{1998PhRvL..80.5255H}. Similarly, the halo number
density decreases due to the free-streaming neutrino action (right
panel of Fig.~\ref{fig:ps_nc}). Again the suppression is larger for
larger neutrino mass, and for a given $\sum m_\nu$ it is more
pronounced for the heaviest, late forming haloes, since the damping in
power from massive neutrinos shifts the maximum cluster mass down
(i.e. the scale beyond which the halo mass function is exponentially
suppressed).

To actually detect the signature of massive neutrinos in the matter
power spectrum and halo mass function we need a cosmic tracer of such
quantities.  As mentioned in the introduction we will use the cluster
number counts and cluster power spectrum as observables to detect
massive neutrino signature. In the next section we will describe the
formalism we use to quantify such an effect on the two observable and
outline the forecasting procedure adopted.

\section{Methodology}\label{sec:meth}
We focus on the galaxy cluster sample which will be provided by a
Euclid-like photometric survey.  For this observable we quantify its
clustering statistics in terms of the average cluster power spectrum,
and its mass distribution in terms of the cluster number
counts. In describing these two quantities we use the same notation
as in ~\cite{sartoris:2011}. In order to extract the cosmological
parameter errors from the mock Euclid data, we perform a Bayesian
likelihood analysis.
 
\subsection{Cluster number counts}\label{sub:nc}
The number of clusters expected for a survey having sky coverage
$\Delta \Omega$ with an observed mass between $M_{\text{l,m}}^{ob}$
and $M_{\text{l,m+1}}^{ob}$ and redshift between $z_\text{l}$ and
$z_{\text{l+1}}$ can be expressed as:
\begin{equation}
 N_{l,m} =  \Delta\Omega \int_{z_\text{l}}^{z_{\text{l+1}}}dz\, \frac{dV}{ dz} \int_{M^{ob}_{\text{l,m}}}^{M^{ob}_{\text{l,m+1}}} \frac{dM^{ob}}{M^{ob}}
        \times \int_0^\infty dM \,n(M,z)\,p(M^{ob}\|M)\,,
\label{eq:nc}
\end{equation}
where $dV/dz$ is the comoving volume element per unit redshift and
solid angle, and $n(M,z)$ is the halo mass function. In this notation
$M_{\text{l,m}=0}^{ob}=M_{\text{min}}(z)$ represents the minimum value
of the observed mass for a cluster to be included in the survey, and
it is determined by the survey selection function
(see \S~\ref{sub:cha}). The integral over the observed mass is computed
within bins having width $\Delta \log M = 0.2$, extending from
$M_{\text{min}}(z)$ to $10^{15.8} h^{-1} M_\odot$. For $n(M,z)$ we
adopt the expression provided by~\cite{2008ApJ...688..709T}, with
mass function parameters obtained for overdensity $\Delta=200$ with
respect to the mean density of the universe (see their Table
2). Moreover, to take into account massive neutrino effects, we follow
the prescription used by many authors~\citep[e.g.][]{2010JCAP...09..014B,2011MNRAS.tmp.1424M,villa12}
neglecting the weakly clustering neutrino component when calculating
the halo mass ($M_\text{halo}=4 \pi r^3 \rho /3$, with
$\rho=\rho_{\rm m}-\rho_\nu$). Many other calibrations of the halo mass
function from simulations have been presented by several
authors~\citep[e.g.][]{1999MNRAS.308..119S, 2001MNRAS.321..372J,
2006ApJ...646..881W, 2010MNRAS.403.1353C}. However, for the purposes
of this work the choice of the best-calibrated mass function has a
minor impact. Indeed the forecast errors depend primarily on the
number of cluster expected for a given cosmological model, which is
far more sensitive to the exponential shape of the mass function rather
than to the calibration details of the mass function.  The factor $p(M^{ob}\|M)$ takes
into account the uncertainties that a scaling relation
introduces in the knowledge of the true cluster mass. Following the
prescription of~\cite{2005PhRvD..72d3006L}, $p(M^{ob}\|M)$ gives the
probability of assigning to a cluster of true mass $M$ an observed mass
$M^{ob}$, as inferred from a giving scaling relation. Under the
assumption of a lognormal-distributed intrinsic scatter around the
nominal scaling relation with variance $\sigma^2_{\ln M}$, the
probability can be written as:
\begin{equation}
p(M^{ob}\|M)\,=\,\frac{\exp[-x^2(M^{ob})]}{ \sqrt{\left( 2\pi \sigma^2_{\ln M}\right) }}\,,
\label{eq:prob}
\end{equation}
where
\begin{equation}
x(M^{ob})\, =\, \frac{\ln M^{ob}-B_{\rm M}-\ln M}{\sqrt{\left( 2 \sigma^2_{\ln M}\right) }}\,.
\label{eq:x_mo}
\end{equation}
Here the parameter $B_{\rm M}$ represents the fractional value of the
systematic bias in the mass estimate. Moreover, according
to~\cite{2010MNRAS.407.2339S} we assume the following parameterization
for redshift dependencies of the halo mass bias and variance (we do
not consider a possible mass dependence of these parameters):
\begin{gather}
B_{\rm M}(z)  =  B_{\rm M,0}(1+z)^\alpha \notag \\
\sigma_{\ln M}(z)  =  \sigma_{\ln M,0}(1+z)^\beta \,.
\label{eq:nuis}
\end{gather}
In our formalism, we have four nuisance parameters, $B_{\rm M,0}$, $\sigma_{\ln
M,0}$, $\alpha$ and $\beta$, which can be allowed to vary
along with the other cosmological parameters during the forecast
procedure (see \S~\ref{ssub:nuis}).

Including Eq.~\ref{eq:prob} into Eq.~\ref{eq:nc} it follows that:
\begin{equation}
N_{\rm l,m}  =  \frac{\Delta\Omega } {2}\int_{z_\text{l}}^{z_{{\rm l}+1}}dz\,\frac{dV} {dz} \int_0^\infty dM\, n(M,z) 
        \times  \left[{\rm erfc}(x_\text{l,m})-{\rm erfc}(x_{\text{l,m}+1}) \right]\, ,
\label{eq:nc2}
\end{equation}
where ${\rm erfc}(x)$ is the complementary error function.

For Euclid data, the photometric redshift measurements will be
  calibrated using a combination of the spectroscopic survey and
  ground-based visual bands photometry, with an expected limiting
  precision of $\sigma(z) \sim 0.05(1+z)$~\citep{2011arXiv1110.3193L}.
  While this error refers to the photometric redshift of a single
  galaxy, the redshift of a cluster identified in the photometric
  survey should be reduced by a factor $N^{1/2}$, where $N$ is the
  number of galaxies assigned to the cluster.  For a typical cluster
  at $z \sim 1.0$ with $N\sim 100$ detected galaxies the error is
  reduced by a factor 10, leading to $\sigma(z) \sim 0.01$.  Thus in
  the following, we assume that the errors on cluster redshift
  measurements can be neglected.

An additional effect on the number counts is induced by line-of-sight
peculiar velocities, which can scatter redshifts by $\delta z \sim
0.003$ for velocities of $\sim 1000\,\text{km s}^{-1}$.  However,
since the redshift bins adopted in the analysis have a width of
$\Delta z = 0.2$, i.e.  far larger than the $\delta z$ value
associated to the peculiar velocities, it is a fair assumption to
neglect this effect.

\subsection{Cluster power spectrum}\label{sub:ps}
In order to include information from the clustering of galaxy
clusters, we calculate the averaged cluster power spectrum within a
given redshift interval using the expression:
\begin{equation}
\bar{P}^{cl}_{\text{l}}(k,z) = \frac{\int^{z_{\text{l+1}}}_{z_l} dz \,\frac{dV}{dz} \,n^2(z)\, P^{cl}(k,z)}{\int^{z_{\text{l+1}}}_{z_l} dz \,\frac{dV}{dz}\, n^2(z) }\,,
\label{eq:barpk}
\end{equation}
 where $n(z)=\int_0^\infty dM n(M,z) \times {\rm
 erfc}(x_{\text{l,m}=0})$ is the comoving number density of clusters
 that are included in the survey at redshift
 $z$~\citep[e.g.][]{2004ApJ...613...41M}. The cluster power spectrum
 $P^{cl}(k,z)$ is expressed in terms of the underlying matter power
 spectrum $P(k,z)$ according to $P^{cl}(k,z)=b_{\eff}^2(z)\,P(k,z)$;
 the term of proportionality $b_{eff}$ is the cluster mass function
 averaged linear bias, defined as:
\begin{equation}
  b_{\eff}(z) = \frac{\int_0^{\infty} dM n(M,z)\,{\rm erfc}(x_{\text{l,m=}0})\, b(M,z)} {\int_0^{\infty} dM\,n(M,z)\,{\rm erfc}(x_{\text{l,m=}0})}\,. 
\label{eq:beff}
\end{equation}
For the linear bias of dark matter halos, $b(M,z)$, we adopt the
fitting function of~\cite{2010ApJ...724..878T} for overdensity
$\Delta = 200$ (see their Table 2). The linear matter power spectrum
$P(k, z)$ is computed with the publicly available software package
{\tt CAMB}~\citep{2000ApJ...538..473L}, which takes correctly into
account the effect of massive neutrinos also in a mild non-linear
regime \citep{bird12}.

As for the effect of errors in photometric redshifts, they are
  expected to introduce a smearing in the power spectrum at small
  scales (see, e.g., \citep{2010MNRAS.401.2477H}), thus degrading the
  information carried by clustering analysis. 
  Consistently with the number counts analysis we neglect in the
  following the effect of uncertainties in redshift
  measurements, thereby not accounting for the damping of the power
  spectrum due to photometric redshift errors. In order to avoid
contribution to the matter power spectrum and scale-dependent bias
introduced by non-linearities, we do not include in our analysis modes
with wavenumbers larger than $k_{\text{max}}=0.1
\,\text{Mpc}^{-1}$~\citep{2009MNRAS.393..297P}. Although massive
free-streaming neutrinos mainly affects the power spectrum at small
scales (see section~\ref{sec:mnu}), a value of $k_\text{max}$ larger
than $\sim 0.3\,\text{Mpc}^{-1}$ would not increase significantly the
sensitivity of the survey. Indeed, given the level of Poisson
\begin{figure}
\centering
\includegraphics[width=10cm]{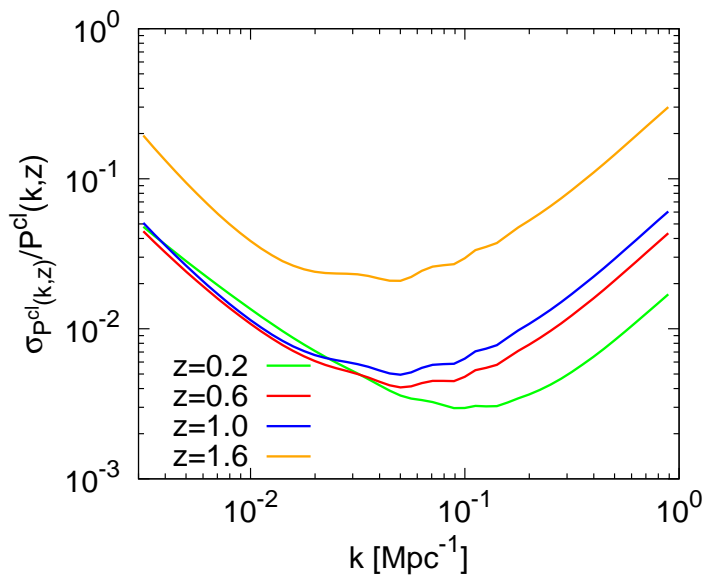}
\caption{Wavenumber dependence of the relative error of the cluster power spectrum, $\sigma_{\bar{P}^{cl}}/\bar{P}^{cl}$ defined as in Eq.~\ref{eqn:sigp}, at four different redshift: $z=0.2, 0.6, 1.0, 1.6$, from bottom to top curves, respectively.}
\label{fig:err}
\end{figure}
noise associated to the cluster distribution (see Eq.~\ref{eqn:sigp},
and Fig.~\ref{fig:err}), higher frequency modes are not adequately
sampled and, therefore, adding them to the analysis does not add
significant information~\citep{2010MNRAS.407.2339S}. As for the
minimum value of the wavenumber we impose
$k_\text{min}=0.003 \,\text{Mpc}^{-1}$; again, using a smaller value of
$k_\text{min}$ does not change the final results, since extremely
large scales are not sampled by the surveys. Finally, in our analysis
we neglect the correction to the power spectrum due to redshift space
distortion effects.

\subsection{Forecasting}\label{sub:fore}
The forecast is based on the Bayesian inference technique, for which
 a likelihood function of the mock data is first constructed and then
 sampled in order to estimate the marginalized probability
 distribution of the parameters. To explore the parameter space by
 means of Monte Carlo Markov Chains we use the publicly available code
 {\tt
 CosmoMC}\footnote{http://cosmologist.info/cosmomc/}~\citep{2002PhRvD..66j3511L},
 where we included a module for the calculation of the cluster number
 counts and power spectrum likelihoods.

Our most general parameter space is:
\begin{equation}
 {\bm \Theta} \equiv (\omega_{\rm b}, \omega_{\rm c}, \Theta_{\rm s}, \tau, n_{\rm s}, \log[10^{10} A_{\rm s}], f_\nu\\
  N_{\eff}, \Omega_{\rm k} ,w , B_{\rm M,0}, \sigma_{\ln  M,0}, \alpha, \beta) \,
\end{equation}
where the first six parameter, which define the standard $\Lambda$CDM
model, are: the physical baryon $\omega_{\rm b}=\Omega_{\rm b} h^2$ and cold dark
matter $\omega_{\rm c}= \Omega_{\rm c} h^2$ densities, the ratio (multiplied by
$100$) between the sound horizon ant the angular diameter distance at
decoupling $\Theta_{\rm s}$, the reionization optical depth $\tau$, the
scalar spectral index $n_{\rm s}$ and the amplitude of initial power
spectrum $A_{\rm s}$. Besides these parameters we performed several
forecasts for different extensions of the minimal cosmological model,
by fitting (along with the other parameters): the neutrino density
fraction $f_\nu=\Omega_\nu/\Omega_{\rm c}$ , the effective number of
neutrino species $N_{\eff}$, the spatial curvature $\Omega_{\rm k}$ and the dark
energy equation of state parameter $w$. Finally, in order to assess
the effect of the uncertain knowledge of the mass-observable relation
we consider also the case in which the four nuisance parameter are
treated as fitting parameters to be determined along with the
cosmological ones.

\begin{table}
\centering{
\caption{Fiducial parameter values.}
\begin{tabular}{ccccccc}
\hline
\hline
$\omega_{\rm b}$ & $\omega_{\rm c}$ & $\Theta_{\rm s}$ & $\tau$ & $n_{\rm s}$ & $\log[10^{10} A_{\rm s}]$ & $ f_\nu$\\ 
\hline
0.02253 & 0.1122 &  1.0395 & 0.085 & 0.967 & 3.18 & 0 \\
\hline
\vspace{0.5ex}\\
\hline
\hline
$N_{\eff}$ & $ \Omega_{\rm k} $ & $w $ & $ B_{\rm M,0}$ & $ \sigma_{\ln  M,0}$ & $ \alpha$ & $ \beta $\\
\hline
 3.046 & 0 & -1 & 0 & 0.45 & 0 & 0 \\
\hline
\end{tabular}
\label{tab:valu}
}
\end{table}

Throughout this paper, our reference model is chosen to be a flat
$\Lambda$CDM model with three neutrino species. The fiducial
$\Lambda$CDM parameter values are listed in Table~\ref{tab:valu},
consistently with the WMAP-7+BAO+$H_0$ best-fit model
by~\cite{2011ApJS..192...18K}. These fiducial parameter values are 
also consistent with the latest WMAP 9-yr best-fit model~\citep{2012arXiv1212.5226H}.

As for the nuisance parameters, clusters mass within Euclid survey will
 be estimated using photometric richness as a cluster mass proxy. 
An accurate calibration of the scaling relation between
   richness and mass will be provided by weak lensing mass
   measurements within the Euclid survey.  We note that different
   authors calibrated the presence of a possible bias in weak lensing
   mass measurements by resorting to cosmological simulations of
   galaxy clusters \cite{2011ApJ...740...25B,2012NJPh...14e5018R}. The
   results of these analyses converge to indicate that a small, but
   sizeable, underestimate in weak lensing masses, is induced by
   projection effects and amounts to 5--10\%. For the purpose of the
   present analysis we prefer to assume that weak lensing provides an
   unbiased calibration of the mass--richness relation, thus fixing
   $B_{\rm M,0}=0$ as a reference value for the mass bias. For the
 intrinsic scatter we assume $\sigma_{\ln M,0}=0.45$ as estimated
 by~\cite{2009ApJ...699..768R} by demanding consistency between
 available weak lensing and X-ray measurements of the maxBCG clusters,
 and the X-ray luminosity-mass relation inferred from the 400d X-ray
 cluster survey.  The intrinsic scatter has the effect of increasing
 the number of clusters included in the survey. Indeed, the number of
 low-mass clusters that are up-scattered above the survey mass limit
 is always larger than the number of rarer high-mass clusters which
 are down-scattered below the same mass
 limit~\citep[e.g.][]{2009PhRvD..79f3009C}.  Because so far there are
 no evidences for the evolution of the nuisance parameters we adopt
 $\alpha=0$ and $\beta=0$ as reference values, thus making the minimal
 assumption of constant bias and scatter with redshift. When the
 nuisance parameters are left free we consider two cases: one with
 strong prior on the evolution parameters, with $\alpha$ and $\beta$
 not allowed to vary with respect to their reference value, and the
 other one with no prior knowledge of their value. The latter turns
 out to be a conservative assumption in view of the large number of
 clusters for which mass measurements from weak lensing will be
 available from Euclid~\citep{2011arXiv1110.3193L}. As such, the
 corresponding uncertainties expected on cosmological parameters
 should be regarded as upper limits of the error introduced by the
 uncertainties in the relation between cluster richness and mass. The
 cluster power spectrum and number counts of the mock data are assumed
 to be equal to the theoretical cluster power spectrum and number
 counts of the fiducial model.

Since we are interested only in parameter error estimation, we define
our likelihood functions $\mathcal{L}$ of the observable $O$ as
\begin{equation}
\chi_{\eff}^2 \equiv -2 \ln \mathcal{L} = \sum_{i,j} \frac{O^{\text{obs}}_{ij} - O^{\text{th}}_{ij}}{\sigma_{O_{ij}}^2}\, ,
\label{eqn:like} 
\end{equation}
in such a way that $\chi_{\eff}^2$ is equal to zero for the fiducial
parameter values. In the previous equation $O_{ij}^{\text{obs}}$
denotes the observed cluster power spectrum,
$\bar{P}_{cl}^{\text{obs}}(k_i,z_j)$ (number counts,
$N^{\text{obs}}(M_i,z_j)$), while $O_{ij}^{\text{th}}$ is the
theoretical cluster power spectrum (number counts) of
Eq.~\ref{eq:barpk} (Eq.~\ref{eq:nc2}). The statistical error
associated to the observed galaxy cluster power spectrum in a bin
centered on $(k_i,z_j)$ ~\citep{1994ApJ...426...23F} reads:
\begin{equation}
 \sigma_{P_{ij}^{cl}}^2 = \frac{(2 \pi)^2 (\bar{P}_{cl}^{\text{th}}(k_i,z_j))^2}{V_{\text{sur}}(z_j) k_i^2 \Delta k} \left[ 1+ \frac{1}{n(z_j)\bar{P}_{cl}^{\text{th}}(k_i,z_j)}\right]^2 \,,
\label{eqn:sigp}
\end{equation}
where $V_\text{sur}(z_j)$ is the comoving survey volume within the
redshift bin centred on $z_j$, and $\Delta k$ is
the size of the bins in wavenumber space. In this way, constraints at
redshift $z$ are mostly contributed by wave-numbers $k$, which maximize
the product $n(z)P_{cl}^{\text{obs}}(k, z)$ (see
Fig.~\ref{fig:err}). The average cluster power spectrum is computed by
integrating over redshift intervals having constant width $\Delta z =
0.2$. This choice of binning represents a compromise between the need
of extracting the maximum amount of information from clustering
evolution and request of negligible covariance between adjacent
z-intervals~\citep[e.g.][]{2010MNRAS.404..239S}. Indeed, the
definition of Eq.~\ref{eqn:like} holds only if the contribution from
different redshift slices carry statistically independent
information. As for the statistical error of the observed number
counts for a given mass and redshift bin centered in $(M_i,z_j)$, we
consider only the Poissonian noise, $\sigma_{N_{ij}^{cl}}^2 =
N^\text{th} (M_i,z_j)$, neglecting the contribution from sample
variance, which accounts for the clustering of clusters due to large scale structure.
Given the large volume to be probed by the Euclid survey
  ($\sim 100 h^{-3}\,{\rm Gpc}$) and the exponential suppression of
  cluster number density for mass larger than the maximum cluster
  mass, the shot-noise errors dominate over sample variance for the
  most of mass and redshift bins~\citep[][also for a more rigorous
  definition of the number counts covariance
  matrix]{2003ApJ...584..702H}. Finally, since number counts
  and power spectrum probe the same mass density field, the covariance
  between the two is expected to be different from zero.  In practice,
  it has been shown \citep[e.g.][]{2007PhRvD..75d3010F} that these two
  observables have in fact negligible covariance.
   
Because the full parameter space is quite large and some parameters
are poorly constrained by LSS observations, we perform our forecast
combining the Euclid-like cluster catalog with Planck-like
data. The mock CMB $TT$, $EE$
and $TE$ power spectra have been simulated following the procedure
of~\cite{2006JCAP...10..013P} according the specifications presented
in the Planck \textit{Blue-Book}~\citep[][page 4, Table
1.1]{2006astro.ph..4069T} based on $14$ months of observations, using
the three frequency channels with the lowest foreground levels at
$100 \text{GHz}$, $143 \text{GHz}$ and $217 \text{GHz}$, and a sky
fraction of $f_\text{sky}=0.80$. In order to avoid problems with
foreground signal, beam uncertainties, etc., we cut-off the spectra at
$l_\text{max}=2000$.

\subsection{Characteristics of the survey}\label{sub:cha}
Euclid is a Medium Class mission of the ESA Cosmic Vision 2015-2025
programme, planned for lunch in 2019. Thanks to its three imaging and
spectroscopic instruments working in the visible and near-infrared
bands, Euclid will cover 15,000 square degrees of extragalactic sky
with the wide survey, thereby providing high--quality images from
visual imaging for more than a billion galaxies, accurate photometric
redshifts from near-IR imaging photometry (in combination with
ground-based data) for about $2\times 10^8$ galaxies and about
$5\times 10^7$ spectroscopic redshift at $z>0.7$ from near-IR slitless
spectroscopy.

The most efficient method to build the Euclid galaxy cluster catalog
relies on the analysis of the photometric data.  To predict
cosmological constraints from the expected sample of galaxy clusters,
we use the analytic selection function adopted in the Euclid
\textit{Red-Book}~\citep{2011arXiv1110.3193L} to forecast the
contribution of the cluster survey to the cosmological
constraints. The computation of the selection function is based on
using the luminosity function of cluster galaxies to compute the
number of galaxies expected within $R_{200c}$\footnote{Here $R_{200c}$
  is defined as the radius encompassing an average density equal to
  200 times the cosmic critical density at a given redshift} down to
the $H_{AB}=24$ magnitude limit reached in the photometric survey, as
a function of the cluster mass and the cluster redshift. 

Specifically, we use an average of the $K_s$-band luminosity functions
of nearby clusters, evaluated within $R_{500c}$ by \cite{LMS03},
which we then evolve passively with redshift \citep{LMGS06}. We
transform the $K_s$ magnitudes into the $H_{AB}$ band by using the
mean color for cluster galaxies. Integrating the luminosity function
down to the apparent magnitude limit of the survey we obtain the
number density of cluster galaxies within $R_{500c}$. Then, after
appropriate scaling and multiplication by the corresponding sphere
volume, we obtain the number of cluster galaxies within a sphere of
radius $R_{200c}$. Given the direct relation between cluster
mass $M_{200c}$ and radius $R_{200c}$, we obtain
the number of observable galaxies for a cluster of given mass
at any redshift.

In practice, this procedure is equivalent to adopting a scaling
relation between a cluster mass $M_{200c}$ and richness, a
relation which evolves with redshift because of passive evolution of
the cluster population, and where the knowledge of the luminosity
function allows the richness to be estimated down to the
redshift-dependent absolute magnitude limits that correspond to the
fixed apparent magnitude limit of the survey.

We then calculate the predicted number of fore-/back-ground galaxies
within a cylinder of angular radius corresponding to $R_{200c}$ at the
cluster redshift, and of length equal to $\pm 3$ times the 
photometric redshift error, with the idea that photometric 
redshifts will be used to reduce the fore-/back-ground. 
The signal-to-noise ($S/N$) for cluster detection is then
 obtained from the ratio between the number of cluster galaxies
 and the rms of the number of fore-/back-ground galaxies. The latter is
contributed by both Poisson noise and cosmic variance.

Assuming $S/N=3$ for the fiducial limiting signal-to-noise
for a reliable cluster detection turns into a selection function which
provides $M_{\rm lim,200c}(z)$ defined as the limiting mass within
$R_{200c}$ for a cluster to be included in the survey. As a result,
one finds $M_{\rm lim,200c}(z)\simeq 1.6\times 10^{14}M_\odot$ at
$z>0.5$, while decreasing at lower redshift, reaching $\simeq 5\times
10^{13}M_\odot$ at $z=0.2$. Finally, to compute Eq.~\ref{eq:nc2} we need to convert
$M_{\rm lim,200c}(z)$ to $M_{\rm lim,200m}(z)$ -- the limiting mass within
 a radius encompassing an overdensity equal to 200 times the mean density
of the Universe -- consistently with the chosen halo mass function (see \S\ref{sub:nc}).
To this end we fallow the recipe given in~\cite{2003ApJ...584..702H},
 assuming a NFW profile~\cite{1997ApJ...490..493N} as halo density profile
and using their fitting formula (C11). 
Even though a $S/N=3$ level may look optimistic the selection
  function adopted in this work is derived using a simplistic
  analytical model which does not take into account any sophisticated
  algorithm for cluster detection, and without making use of the full
  information available (e.g. from cluster density profiles,
  luminosity functions, red sequence and spectroscopic data).
Therefore, the chosen limiting signal-to-noise is likely to
 represent a conservative estimate, and we can assume
the cluster sample to be $100\%$ complete and pure.
Clearly, a detailed assessment of the completeness and purity of the
cluster sample should require a detailed analysis of the performance
of different cluster detection algorithms when applied to the Euclid
survey, which is beyond the aim of this paper. Moreover, as discussed in~\cite{2010ApJ...708..645R},
what matters in parameter estimation is not the level of the survey completeness and purity, 
but the uncertainty in their calibration. Thus, the assumption of a $100\%$ pure
and complete sample for $S/N\geq 3$ can be considered as assuming that purity and completeness
will be accurately measured in this regime.

\begin{figure}
\centering
\includegraphics[width=10cm]{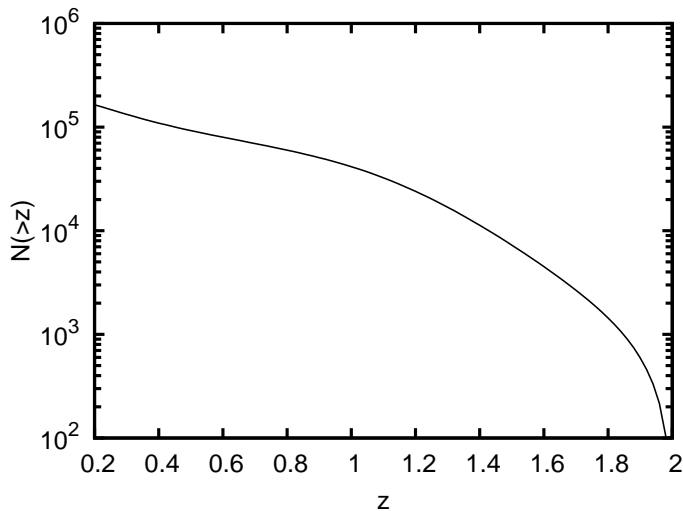}
\caption{The cumulative cluster redshift distribution as predicted by
  the reference cosmological model and the reference values for the
  mass nuisance parameters (see Table 1), for the Euclid cluster
  survey.}
\label{fig:n_z}
\end{figure}

For the sky coverage we adopt the required area for the wide Euclid
survey $\Delta \Omega = 15,000 \text{deg}^2$
\citep{2011arXiv1110.3193L}, while the cluster number counts and power
spectrum are evaluated in redshift bins of width $\Delta z=0.2$
between $z=0.2$ and $z=2$. Moreover, as discussed in
  \S\ref{sub:nc}, the cluster number counts is computed within bins of
  observed mass having width $\Delta \log M = 0.2$ and extending from
  $M_{\text{l,m}=0}^{ob}=M_{\rm lim,200m}(z)$ to $10^{15.8} h^{-1}\,
  M_\odot$. From Eq.~\ref{eq:nc2}, using this survey specifics and
fiducial parameter values, we expect that Euclid will find of order
$1.5 \times 10^5$ cluster with a S/N better than 3, between $z=0.2$
and $z=2.0$, with $\sim 4 \times 10^4$ having $z>1$
(see Fig.~\ref{fig:n_z}).\\

\section{Analysis and results}\label{sec:res}
Having defined the reference cosmological model and the specifics of
Euclid survey we now present forecast errors on neutrino parameters
for various extensions of the minimal $\Lambda$CDM model. For each of
the cases that we describe here below, we run four independent chains,
requiring the fulfillment of the Gelman \& Rubin~\cite{1992...Gelman}
criteria with $R-1\leqslant 0.03$ as convergence test.

\subsection[Three massive neutrinos]{Three massive neutrinos: $\Lambda$CDM+$m_\nu$}\label{ssub:nu}
\begin{figure}
\includegraphics[width=7.3cm]{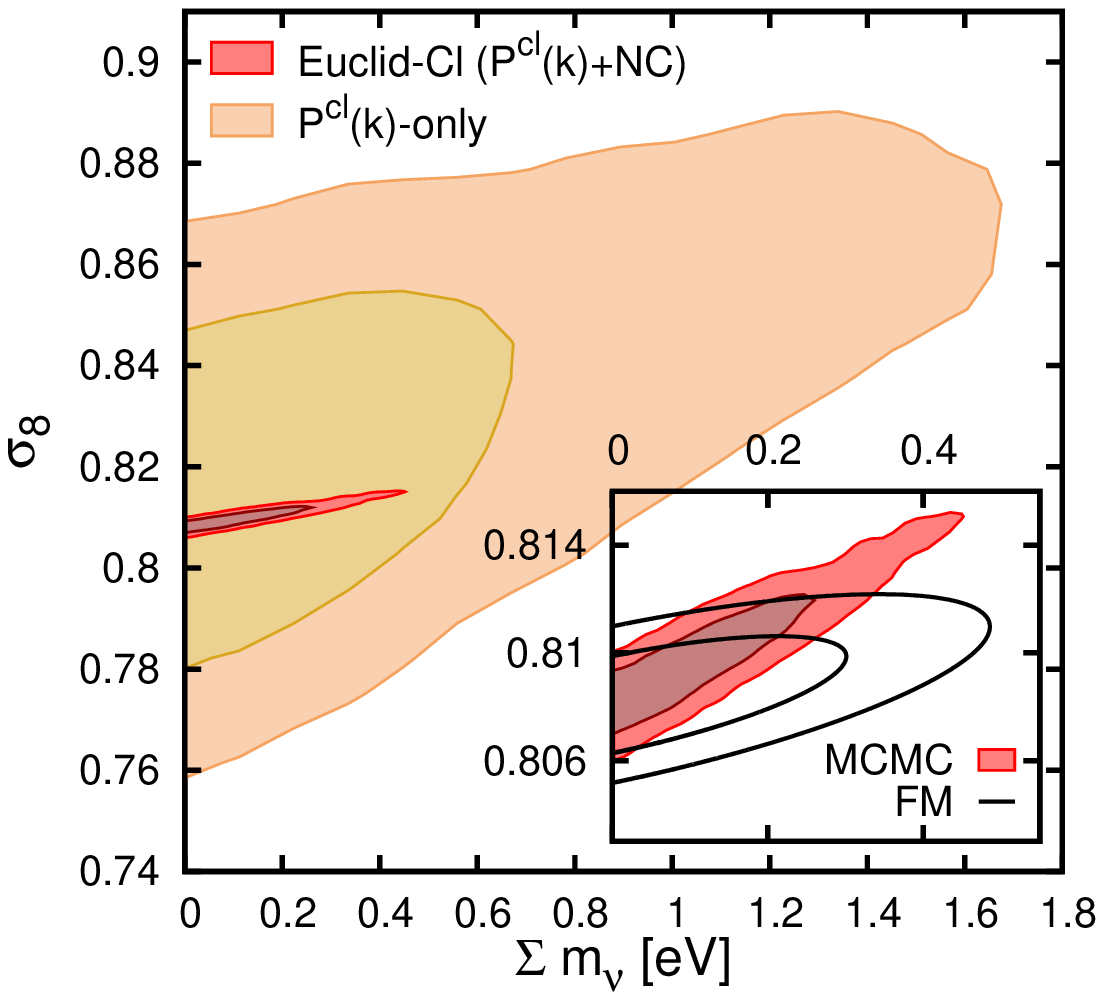}
\includegraphics[width=7.3cm]{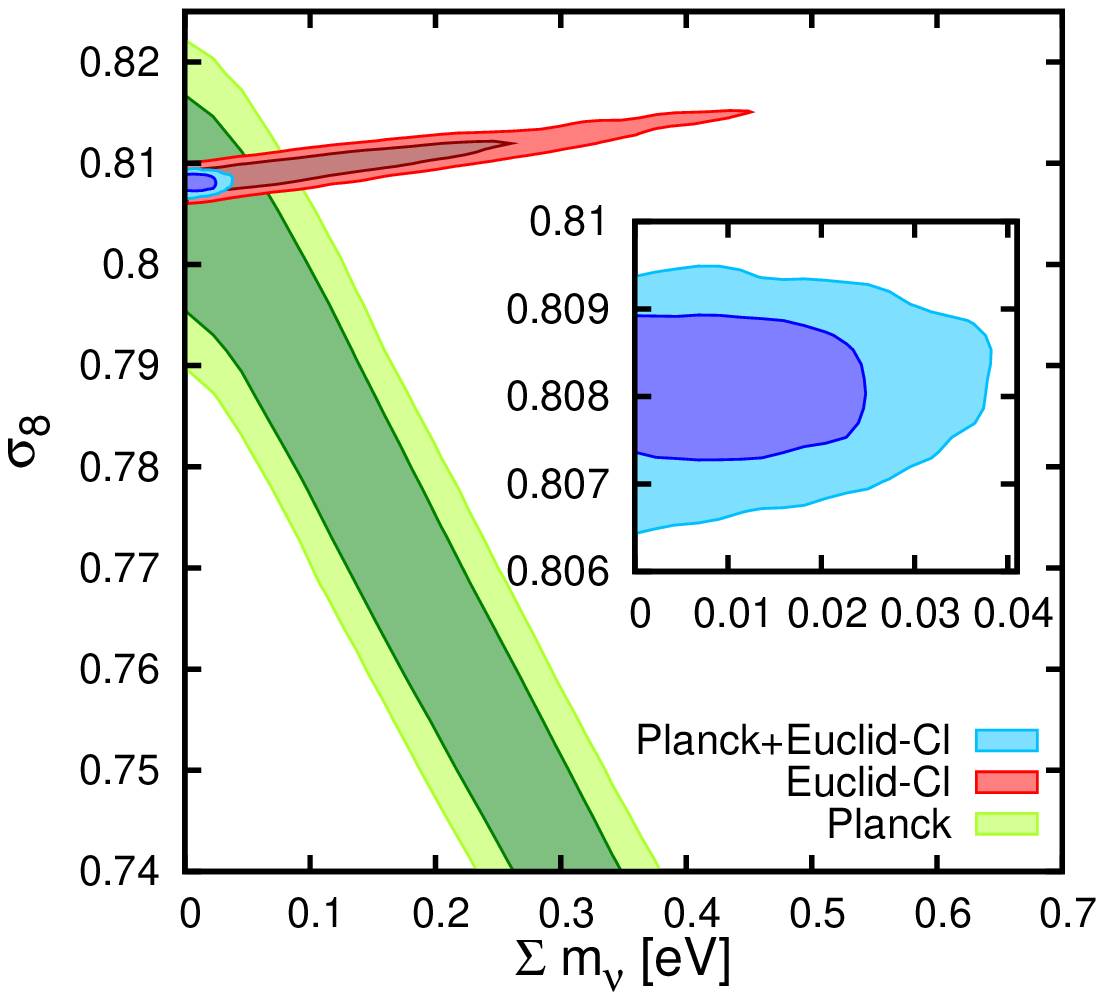}
\caption{The $68\%$ and $95\%$ CL contours in the $(\sum
  m_\nu$-$\sigma_8)$ plane for a $\Lambda$CDM+$m_\nu$ model. Left
  panel: contours from cluster power spectrum (large contours;
  $P^{cl}$-only) and the combination of cluster power spectrum and
  number counts (small contours; Euclid-Cl). The insert plot shows a
  zoom of the confidence contours given by the Euclid-Cl dataset
  compared with the contours obtained from the Fisher Matrix technique
  using the same dataset. Right panel: contours from Planck
  (\textit{green}), Euclid-Cl (\textit{red}) and Planck+Euclid-Cl
  (\textit{blue}) datasets. The insert plot shows a zoom of the
  confidence contours obtained from the Planck+Euclid-Cl datasets.}
\label{fig:mnu_eu}
\end{figure}

\begin{figure}
\includegraphics[width=7.3cm]{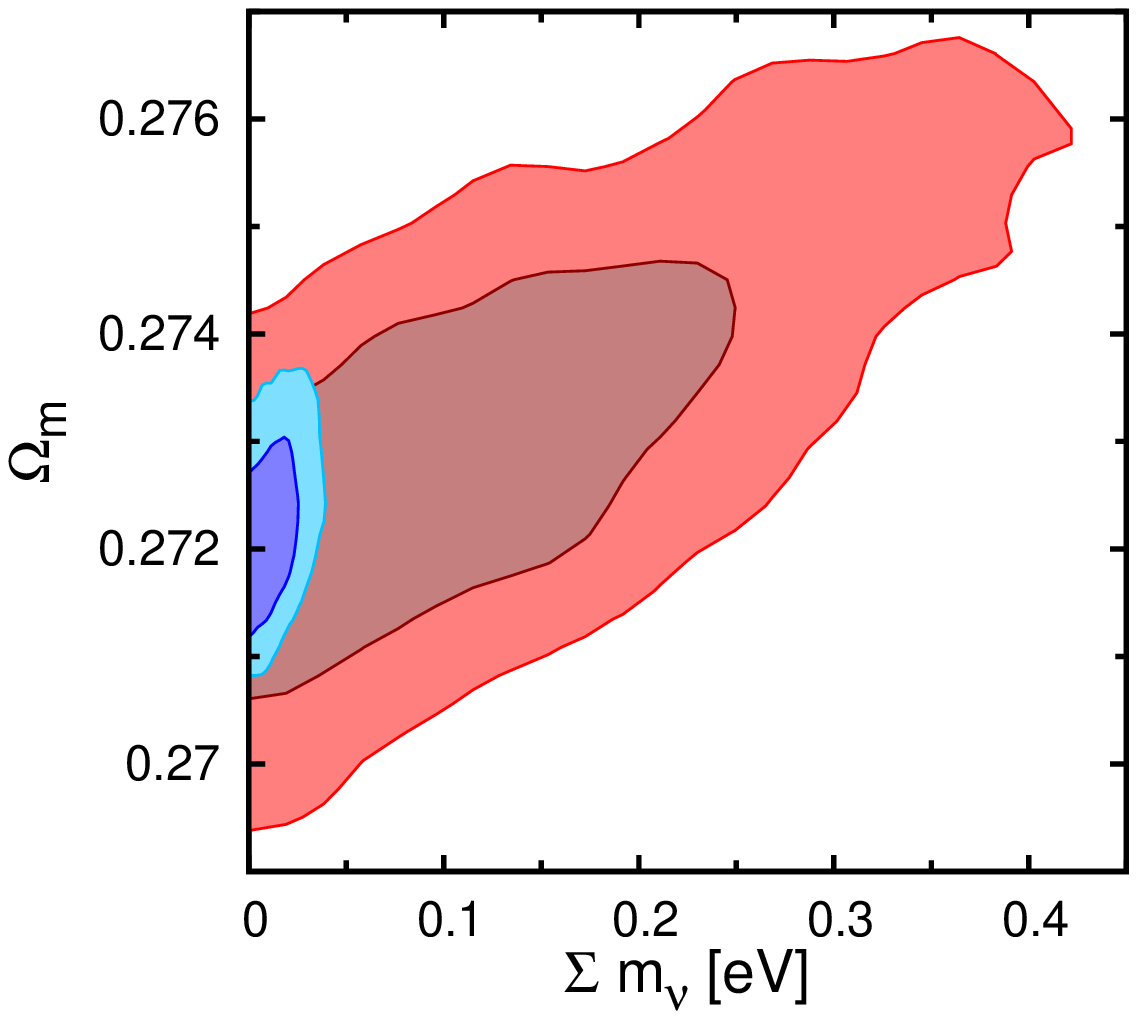}
\includegraphics[width=7.3cm]{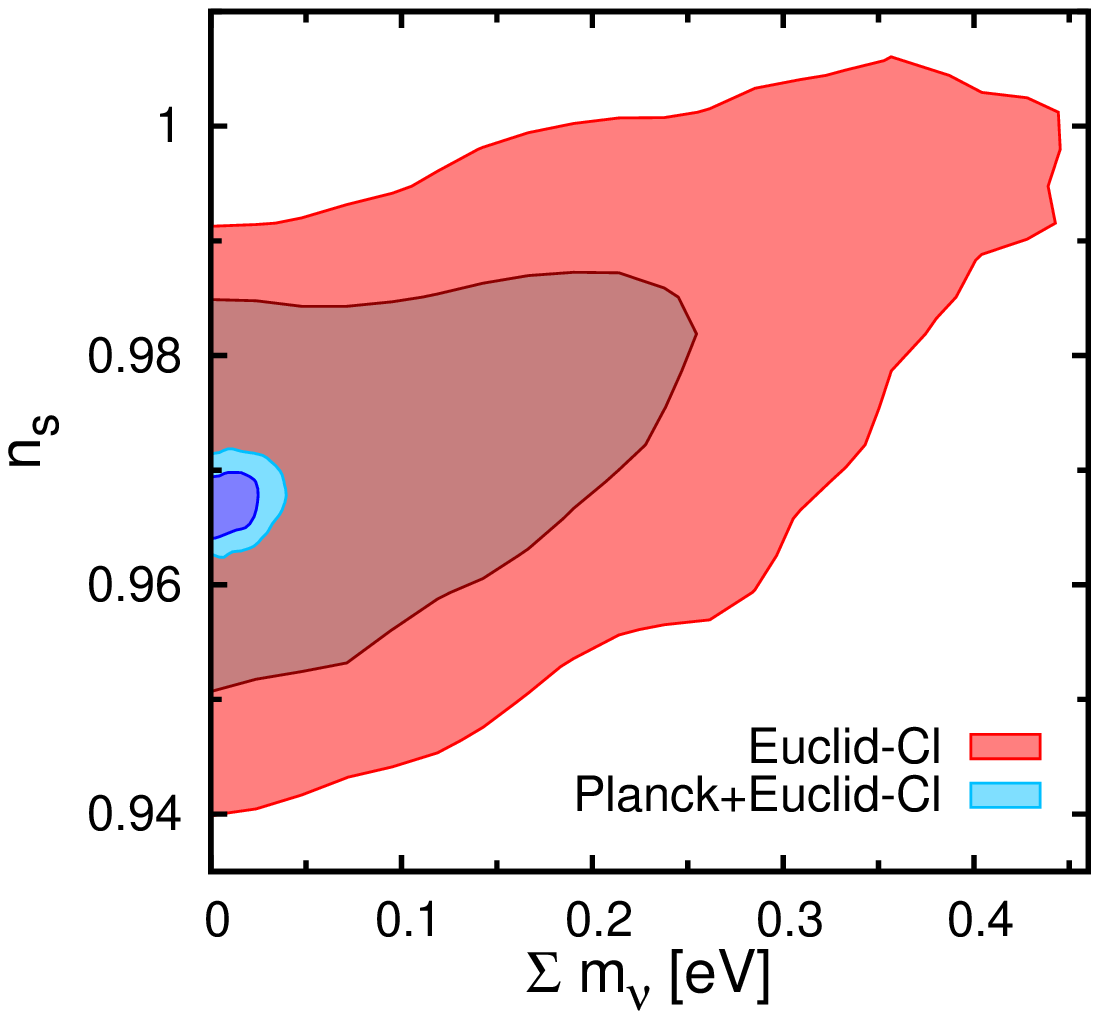}
\caption{The $68\%$ and $95\%$ CL contours in the $\sum
  m_\nu-(\Omega_{\rm m},n_{\rm s})$ planes for a $\Lambda$CDM+$m_\nu$
  model, from Euclid-Cl (red contours) and the Planck+Euclid-Cl (blue
  contours) datasets. When only Euclid-Cl dataset is used the
  parameter $\tau$, which is not constrained by this data, is kept
  fixed to its fiducial value $0.085$.}
\label{fig:mnu}
\end{figure}

We start by considering the scenario with three degenerate massive
neutrino species. In Table~\ref{tab:nu} we report the $68\%$ and $95\%
\,$ CL bounds on $\sum m_\nu$ derived from different data sets: Planck
only, cluster power spectrum only, cluster number counts and power
spectrum (hereafter Euclid-Cl), and the combination of Planck and
Euclid-Cl. When only cluster power spectrum data are considered we
obtain a quite loose $2\sigma$ upper limit on $\sum m_\nu$ of
$1.20\,\text{eV}$.  Otherwise, the information contained in cluster
number counts alone is unable to constrain the total neutrino mass,
but it greatly improves the error on $\sum m_\nu$ once added to
cluster power spectrum data, mainly thanks to the tight constraints
provided on $\sigma_8$ (see left panel of Fig.~\ref{fig:mnu_eu}).
Specifically, the upper limit for $\sum m_\nu$ shrinks by a factor
$\sim 4$ to $0.35\, \text{eV}$ ($95\%$CL). This error is comparable to
the present constraints obtained combining CMB and LSS probes, and of
the same order of magnitude of the error expected for
Planck. Regarding parameter degeneracies for the galaxy cluster
dataset, the total neutrino mass is correlated with all the
cosmological parameter affecting the galaxy power spectrum shape
(i.e. $\Omega_{\rm m}$, $\sigma_8$, $n_{\rm s}$; see red contours in
Fig.~\ref{fig:mnu_eu} and Fig.~\ref{fig:mnu}).
\begin{table}
    \centering
    \caption{Constraints on $\sum m_\nu$ for $\Lambda$CDM+$m_\nu$ model from Planck, cluster power spectrum ($P^{cl}-only$), Euclid-Cl (cluster number counts and power spectrum) data, and the combination of the two data sets Euclid-Cl+Planck. Because the parameter $\tau$ is not constrained by Euclid data, when CMB measurements are not included $\tau$ is kept fixed to its fiducial value.}
    \label{tab:nu}
    \vspace{1mm}
   \begin{tabular}{llcccc}
    \hline 
	Model			&	      &	\multicolumn{4}{c}{$\Lambda$CDM+$m_\nu$} \\
    \hline 
	Data				      &			& Planck	& $P^{cl}-$only	& Euclid-Cl	& Euclid-Cl+Planck\\
    \hline
    \hline
    \multirow{2}{*}{$\sum m_\nu\,[\text{eV}]$} & $68\%$ CL	& $<0.41$	& $<0.41$	& $<0.17$	& $<0.017$	\\
					      & $95\%$ CL	& $<0.74$	& $<1.20$	&$<0.35$	& $<0.031$	\\

   \hline

    \end{tabular}
\end{table}

The main power of constraints in cosmological parameters indeed
originate from the joint analysis of galaxy cluster and CMB
datasets. In this case the error on $\sum m_\nu$ is reduced to
$31 \,\text{meV}$; an improvement of more than one order of magnitude
that would allow a $2\sigma$ detection of the total neutrino mass even
in the minimal normal hierarchy scenario ($\sum m_\nu \simeq
0.05 \, \text{eV}$). The reason for such an improvement can be easily
understood by looking at the right panel of Fig.~\ref{fig:mnu_eu} which shows the $68\%$ and
$95\%$ confidence regions in the $(\sum m_\nu-\sigma_8)$ plane from Planck data, Euclid-Cl data and the
combination of the two. Taken independently, the CMB and galaxy
cluster data exhibit significant degeneracies in this plane, but the
nearly orthogonal degeneracy directions allow their combination to
provide tight constraint on these parameters, and in particular on the
neutrino mass. When Planck priors are added to the Euclid-Cl
constraints, all degeneracies are either resolved or largely
reduced (see blue contours in Fig.~\ref{fig:mnu_eu} and Fig.~\ref{fig:mnu}).
Similar levels of sensibility on $\sum m_\nu$ are also
expected combining Euclid galaxy or cosmic shear power spectrum
measurements with Planck CMB data~\citep[e.g.][]{2011JCAP...03..030C, 2012arXiv1210.2194A,
2012JCAP...11..052H,2008IJMPD..17.2025X} and from the combination of
Planck SZ cluster survey and Planck CMB data~\citep{Mak:2013jia}.

\subsection[Varying the effective number of neutrinos]{Varing $N_{\eff}\,$: $\Lambda$CDM+$m_\nu$+$N_{\eff}$}\label{ssub:neff}
\begin{figure}
\includegraphics[width=7.3cm]{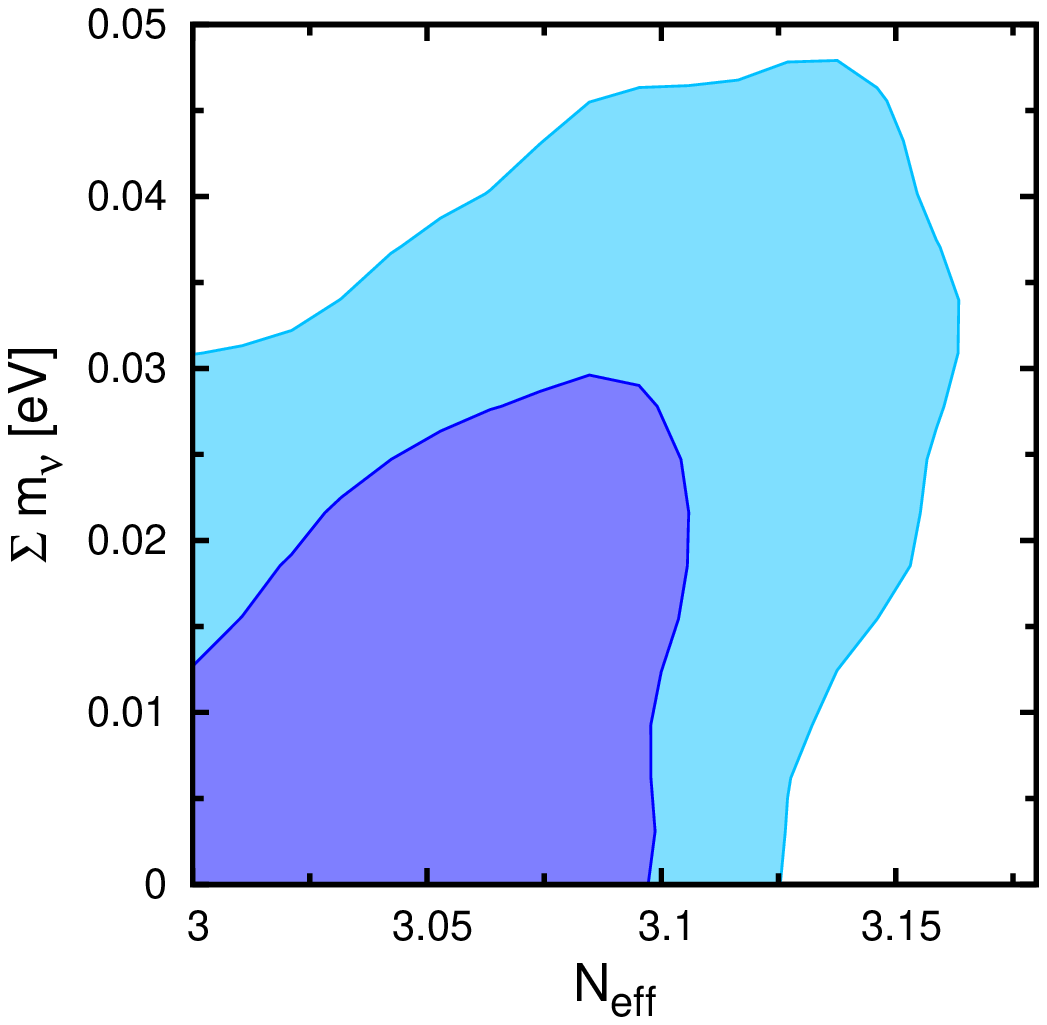}
\includegraphics[width=7.3cm]{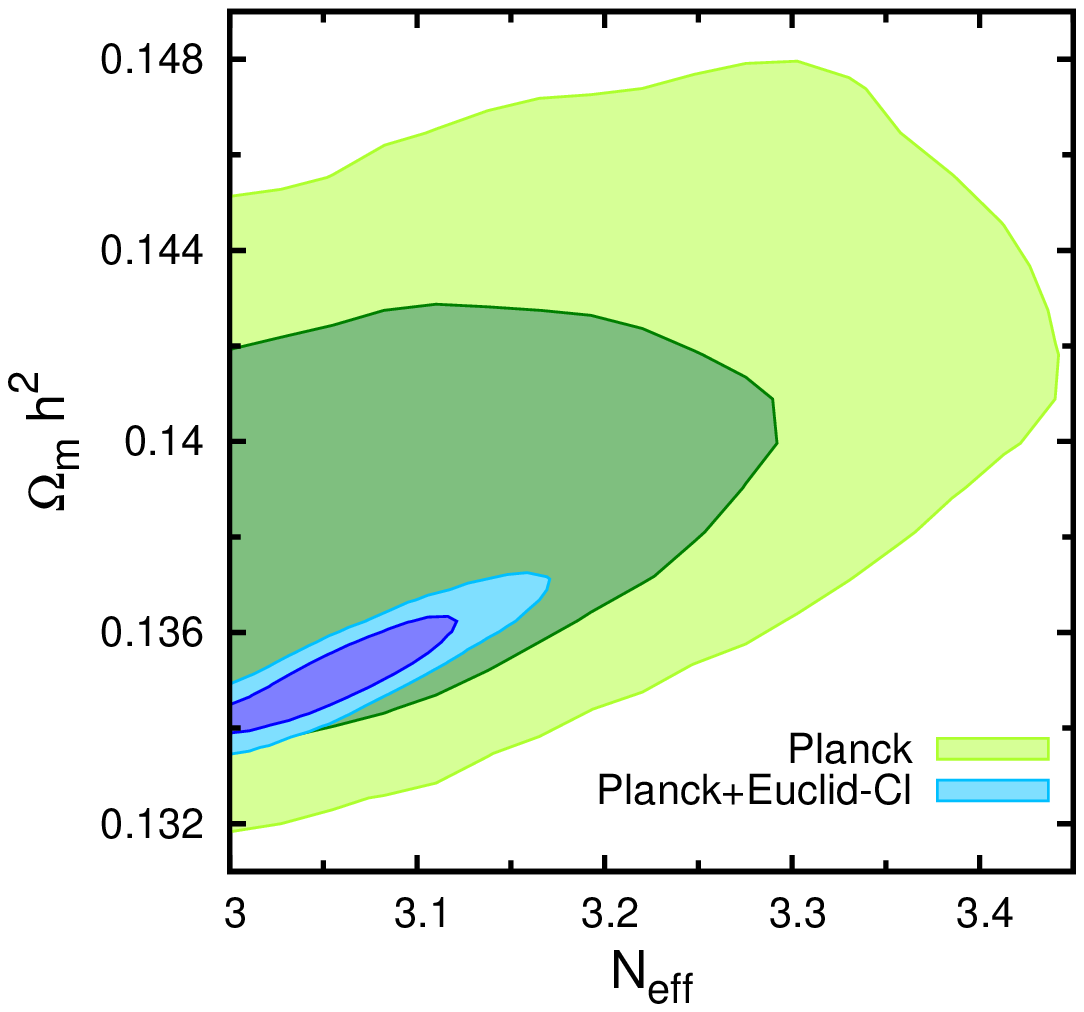}
\caption{The $68\%$ and $95\%$ CL contours in the $N_{\eff}-(\Omega_{\rm m}, \sum m_\nu)$ planes for a $\Lambda$CDM+$m_\nu$+$N_{\eff}$ model, from:  Planck (green contours) and Planck+Euclid-Cl (blue contours) data.}
\label{fig:neff}
\end{figure}

We now explore the scenario with massive neutrinos and $N_{\eff}$
effective number of neutrino species. Again, we distribute the sum of
neutrino masses equally among three active species $(N_\nu=3)$, and we
treat additional contribution to $N_{\eff}$ as massless, such that
$N_{\eff}= 3 + \Delta N_{\eff}$, with the prior $\Delta
N_{\eff}\geqslant 0$.  While the choice of keeping $N_\nu$ fixed does not
affect constraints from CMB measurements (what matters is $N_{\eff}$),
it could change the sensitivity to $\sum m_\nu$ and $N_{\eff}$ based
on galaxy clusters data. Indeed, changing $N_\nu$ would change
the mass of each massive neutrino and thus its free-streaming lenght
beyond which the power spectrum is suppressed (see
Eq.~\ref{eqn:kfs}). We checked the case with fixed $\Delta N_{\eff}$
and $N_\nu$ as free parameter and we find no qualitative changes in
our results. This means that the data are not sensitive to the exact
position of the break in the power spectrum induced by free-streaming
neutrinos, and thus to the neutrino mass hierarchy.
\begin{table}
    \centering{
    \caption{Constraints on $\sum m_\nu$ and $N_{\eff}$ for $\Lambda$CDM+$m_\nu$+$N_{\eff}$ model.}
    \label{tab:mnuneff}
    \vspace{1mm}
   \begin{tabular}{llcc}
        \hline 
	Model				&      &	\multicolumn{2}{c}{$\Lambda$CDM+$m_\nu$+$N_{\eff}$} \\
    \hline 
	Data				      &			& Planck	& Euclid-Cl+Planck	\\
    \hline
    \hline
    \multirow{2}{*}{$\sum m_\nu\,[\text{eV}]$} & $68\%$ CL	& $<0.42$	& $<0.022$	\\
					      & $95\%$ CL	& $<0.78$	& $<0.040$	\vspace{0.5ex} \\
    $N_{\eff}$					& $95\%$ CL	& $<3.36$	& $<3.14$	\\

   \hline

    \end{tabular}
}
\end{table}
Table~\ref{tab:mnuneff} shows the joint constraints on the sum of
neutrino masses and on the effective number of neutrino species from
Planck data alone and the combination of Planck and Euclid-Cl
datasets.  Looking at Planck data alone, the quality of the
constraints on $\sum m_\nu$ are nearly unchanged from the
single-parameter extensions discussed earlier, as it would be expected
for independent parameters. Indeed, $\sum m_\nu$ and $N_{\eff}$ are
constrained by different features in the CMB spectra: the early
integrated Sachs-Wolfe effect for $\sum m_\nu$, the damping scale and
the position of the acoustic peaks for $N_{\eff}$~\citep[see e.g.][and
references therein ]{2012arXiv1212.5226H}.  However, since both
$\sum m_\nu$ and $N_{\eff}$ have similar effects on the matter power
spectrum (see \S~\ref{sec:mnu}), the correlation of the two degrades
the upper bound on the sum of neutrino masses inferred from
Euclid-Cl+Planck data by $\sim 30\%$ to $\sum m_\nu<0.040[\text{eV}]$
at $95\%$CL. With this accuracy, it would still be possible a
$2\,\sigma$ detection of neutrino masses in the minimal normal
hierarchy scenario. Constraining $N_{\eff}$ is mainly achieved through
CMB measurements of the redshift of the matter-radiation equality
$z_{eq}$ and the baryon density $\Omega_{\rm b} h^2$. However, keeping
$z_{eq}$ and $\Omega_{\rm b} h^2$ fixed as $N_{\eff}$ increases can be
achieved by increasing the cold dark matter density $\Omega_{\rm c}
h^2$, which displays a large correlation with
$N_{\eff}$~\citep{2011arXiv1104.2333H}. Euclid-Cl data alone
are unable to provide constraints on the number of effective
species. However, the inclusion of clusters dataset allows to
significantly improve the measurements of $\Omega_{\rm m} h^2$ (by
constraining $\sigma_8$), thus reducing the $2\sigma$ error on the
effective number of neutrino by a factor larger than $2.5$ from $0.36$
to $0.14$ (see right panel of Fig.~\ref{fig:neff}). After the inclusion
of Euclid-Cl data $N_{\eff}$ still exhibits strong degeneracies with
many cosmological parameters (e.g. $\Omega_{\rm m} h^2$, $H_0$ and
$n_{\rm s}$), and a correlation of $\sim0.5$ with $\sum m_\nu$
(see left panel of Fig.~\ref{fig:neff}).
\subsection[Extended models]{Extended models: $w$CDM+$m_\nu$+$N_{\eff}$ and curved Universe}\label{ssub:ext}
\begin{figure}
\includegraphics[width=7.3cm]{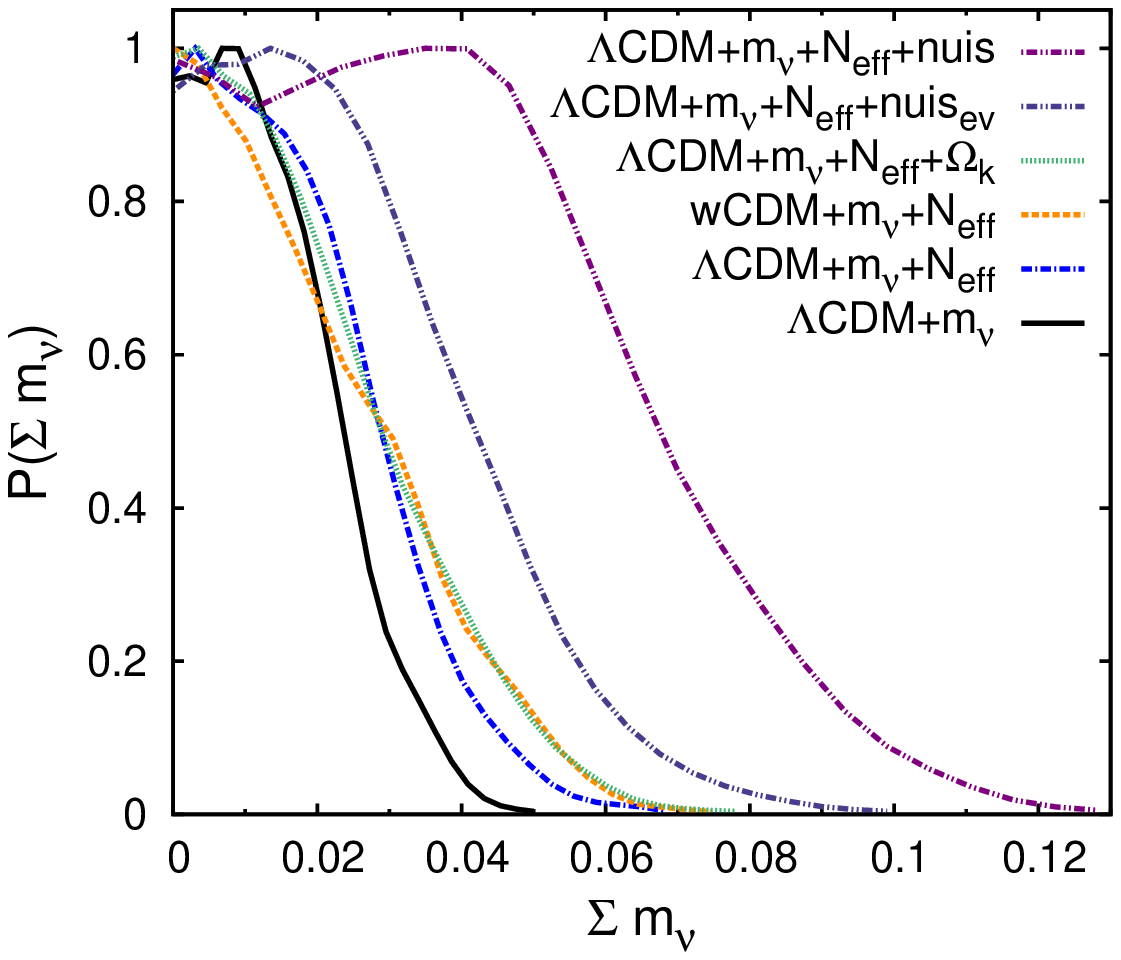}
\includegraphics[width=7.3cm]{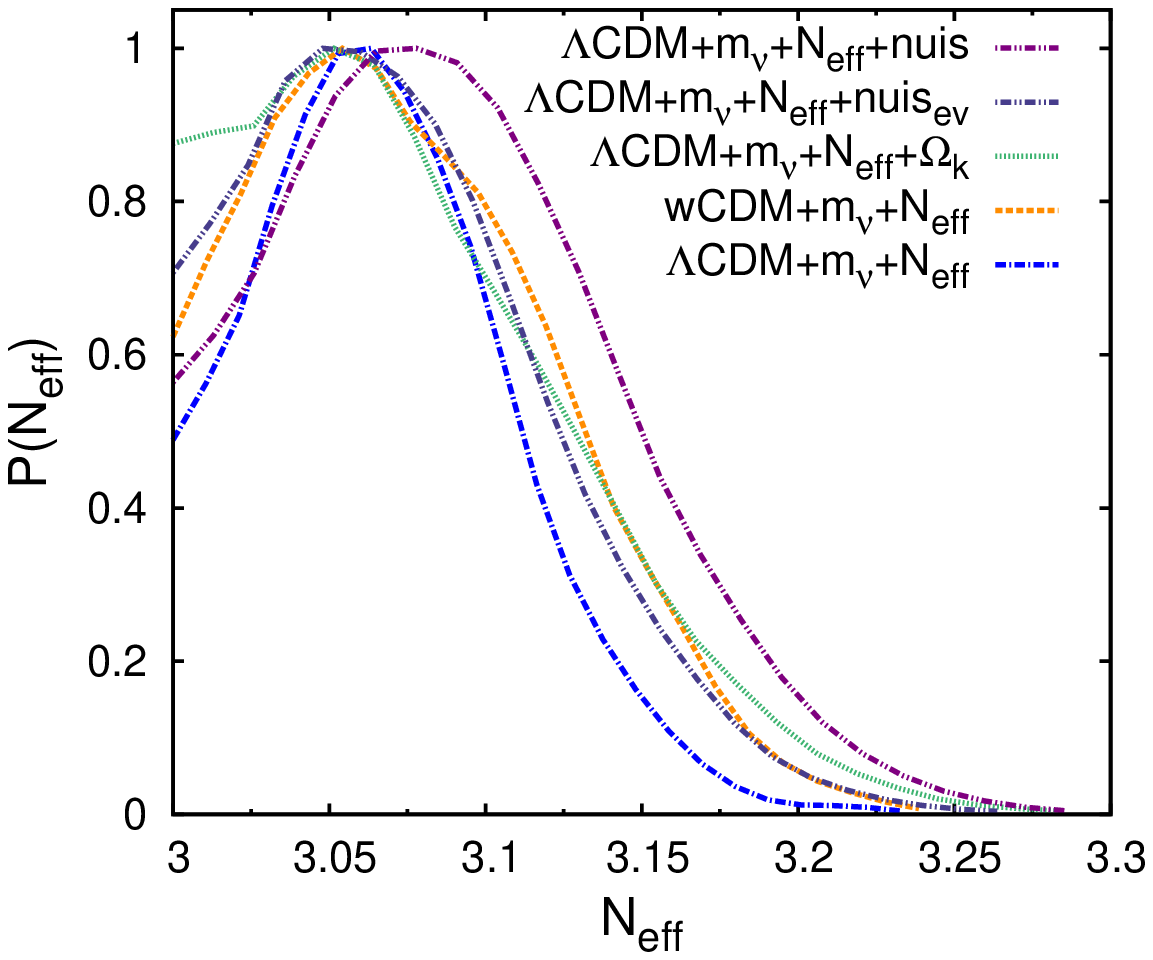}
\caption{The marginalized one-dimensional posteriors for $\sum m_\nu$
  (left) and $N_{\eff}$ (right) for different parameter extensions
  from the combination of Euclid-Cl and Planck datasets.}
\label{fig:1dmnuneff}
\end{figure}

\begin{figure}
\center
\includegraphics[width=10cm]{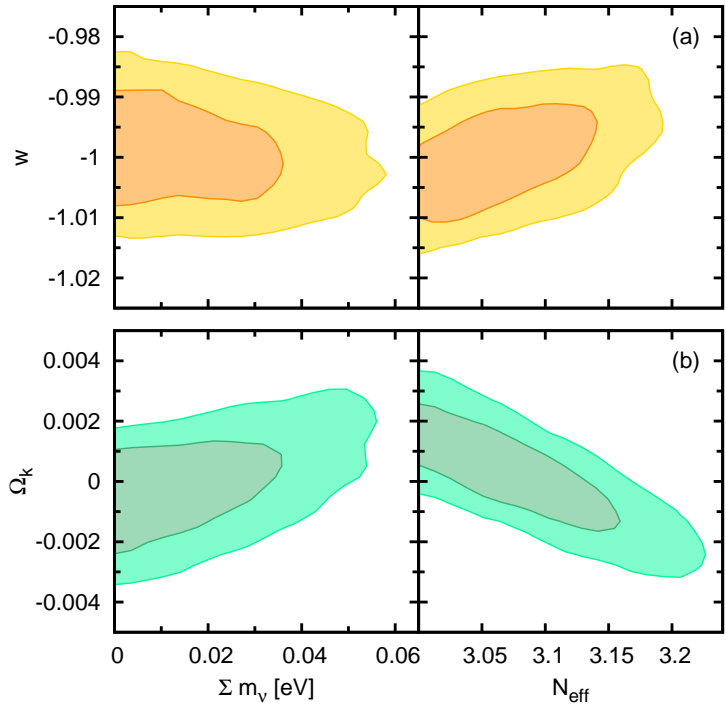}
\caption{Joint two dimensional marginalized constraints on $w$ (a) and
  $\Omega_{\rm k}$ (b) against $(\sum m_\nu, N_{\eff})$ at $68\%$ and
  $95 \%$ CL. The confidence regions are respectively for the extended
  parameter space $w$CDM+$m_\nu$+$N_{\eff}$ and
  $\Lambda$CDM+$m_\nu$+$N_{\eff}+\Omega_{\rm k}$, using data from
  Euclid-Cl+Planck.}
\label{fig:w}
\end{figure}

\begin{table}
    \centering{
    \caption{Constraints on $\sum m_\nu$ and $N_{\eff}$ for the two parameter extensions ($w, \Omega_{\rm k}$) from Euclid-Cl+Planck datasets.}
    \label{tab:float}
    \vspace{1mm}
   \begin{tabular}{llcc}
    \hline
    Data &							& \multicolumn{2}{c}{Planck+Euclid-Cl} \\
    \hline 
	Model				      &			& $w$CDM+$m_\nu$+$N_{\eff}$& $\Lambda$CDM+$m_\nu$+$N_{\eff}$+$\Omega_{\rm k}$ \\
    \hline
    \hline
    \multirow{2}{*}{$\sum m_\nu\,[\text{eV}]$} & $68\%$ CL	& $<0.024$		& $<0.024$								\\
					      & $95\%$ CL	& $<0.046$		& $<0.046$								\vspace{0.5ex} \\
    $N_{\eff}$					& $95\%$ CL	& $<3.16$		& $<3.17 $								\\

   \hline
    \end{tabular}
  }
\end{table}
Next, we consider how the constraints on $\sum m_\nu$ and $N_{\eff}$
are affected when additional degrees of freedom are introduced in the
cosmological model. The effect on $(\sum m_\nu, N_{\eff})$ of adding
these degree of freedom to the $\Lambda$CDM+$m_\nu$+$N_{\eff}$ model
are shown in Fig.~\ref{fig:1dmnuneff} and listed in
Table~\ref{tab:float}.  We start by considering a constant dark energy
equation of state ($w \neq -1$). Neutrino properties ($\sum m_\nu,
N_{\eff}$) and $w$ are generally degenerate because they can both
affect the shape of the matter and CMB power spectra~\citep[e.g.][]{2007PhRvD..75j3505X}.
Looking at Fig.~\ref{fig:w} (a), we indeed see this degeneracy in the plane
$N_{\eff}-w$, which displays a correlation of $\sim 0.5$, whereas the
parameters $w$ and $\sum m_\nu$ show almost no correlation. The Euclid
clusters catalog, probing the evolution of the LSS up to $z \sim 2$,
will be able to put tight constraints on the dark energy equation of
state; we find for the combination of Planck and Euclid-Cl data:
$-1.011<w<-0.987$ ($95 \%$CL). Given the small uncertainty on $w$ the
constraints on neutrino mass and effective number of species are only
slightly degraded when $w$ is allowed to vary; the $95\%$ CL upper
limit for $N_{\eff}$ is relaxed from $3.14$ to $3.16$ due to the
degeneracy with $w$. Whereas, the $95\%$ CL upper limit for $\sum
m_\nu$ undergoes only a small degradation from $0.040\, \text{eV}$ to
$0.046\, \text{eV}$, caused by the weak constraints on parameters that
are correlated with $\sum m_\nu$ induced by the extension of the
parameter space. Secondly, we relax the prior on the curvature of the universe by
considering the case $\Lambda$CDM+$m_\nu$+$N_{\eff}+\Omega_{\rm k}$. Since
current data do not support departures from the flat $\Lambda$CDM
model either through $\Omega_{\rm k} \neq 0$ or $w \neq -1$, we introduce
these parameters separately. From the combination of Planck and
Euclid-Cl datasets we obtain, for the curvature parameter, the
following constraint: $-0.0024<\Omega_{\rm k}<0.0024$ ($95 \%$CL).
As CMB power spectrum suffer from a well known \textquotedblleft
geometrical degeneracy\textquotedblright~\citep[e.g.][]{1997MNRAS.291L..33B, 1997ApJ...488....1Z}, 
Euclid-CL data considerably improves the error on $\Omega_{\rm k}$ breaking
such degeneracy thanks to the tight constraint on $\Omega_{\rm m}$ (given by
the growth information encoded in the dataset).
The spatial curvature mainly affects the expansion rate via the Friedmann
equation, as well as the total neutrino mass and number of effective
species do. As it can be seen in Fig.~\ref{fig:w} (b), this results in a
correlation with both $\sum m_\nu$ and $N_{\eff}$ of the order of
$\sim0.5$ and $\sim0.6$, respectively. Despite these quite large
degeneracies with $\Omega_{\rm k}$, the small error associated to the
curvature parameter leads to a slight relaxation of the constraints on
neutrino properties: the upper limit for neutrino mass degrades by
$\sim 10\%$, passing from $0.040\, \text{eV}$ to $0.046\, \text{eV}$
($95 \%$CL), while the $2\sigma$ error on $N_{\eff}$ shift from $0.14$ to
$0.17$, a $20\%$ degradation.

Thus, in both cases, the parameter extension entails a relaxation of
the constraints on $\sum m_\nu$ and $N_{\eff}$; nonetheless, given the
high accuracy with which $w$ and $\Omega_{\rm k}$ are expected to be
measured, the survey would still allow a $2\sigma$ detection of
neutrino mass in the minimal normal hierarchy scenario and reveal the
presence of possibles extra relativistic species. 
\subsection[Nuisance parameters]{Nuisance parameters}\label{ssub:nuis}
\begin{table}
  
    \centering{
      \caption{Constraints on $\sum m_\nu$ and $N_{\eff}$ for $\Lambda$CDM+$m_\nu$+$N_{\eff}$ models with free nuisance parameters.}
    \label{tab:nuis}
    \vspace{1mm}
   \begin{tabular}{llcc}
\hline
    Data &							& \multicolumn{2}{c}{Planck+Euclid-Cl} \\
    \hline 
	Model				      &			& $\Lambda$CDM+$m_\nu$+$N_{\eff}$+nuis 	& $\Lambda$CDM+$m_\nu$+$N_{\eff}$+$\text{nuis}_{\text{ev}}$ 	\\
    \hline
    \hline
    \multirow{2}{*}{$\sum m_\nu\,[\text{eV}]$} & $68\%$ CL	& $<0.049$					& $<0.031$						\\
					      & $95\%$ CL	& $<0.083$					& $<0.056$				\vspace{0.5ex} \\
    $N_{\eff}$					& $95\%$ CL	& $<3.18$					& $<3.16$						\\

   \hline

    \end{tabular}
  }

\end{table}

\begin{figure}
\includegraphics[width=7.3cm]{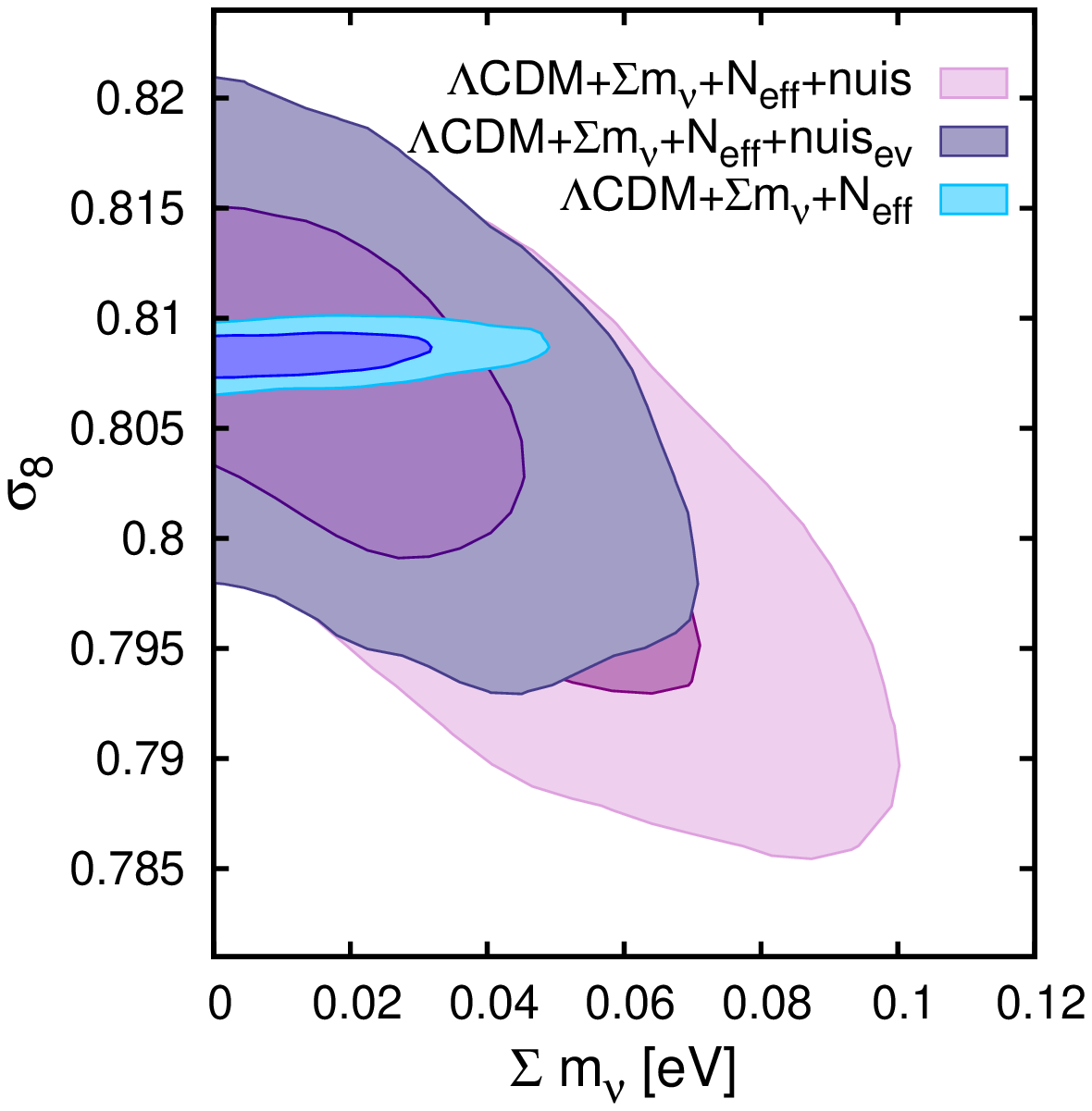}
\includegraphics[width=7.3cm]{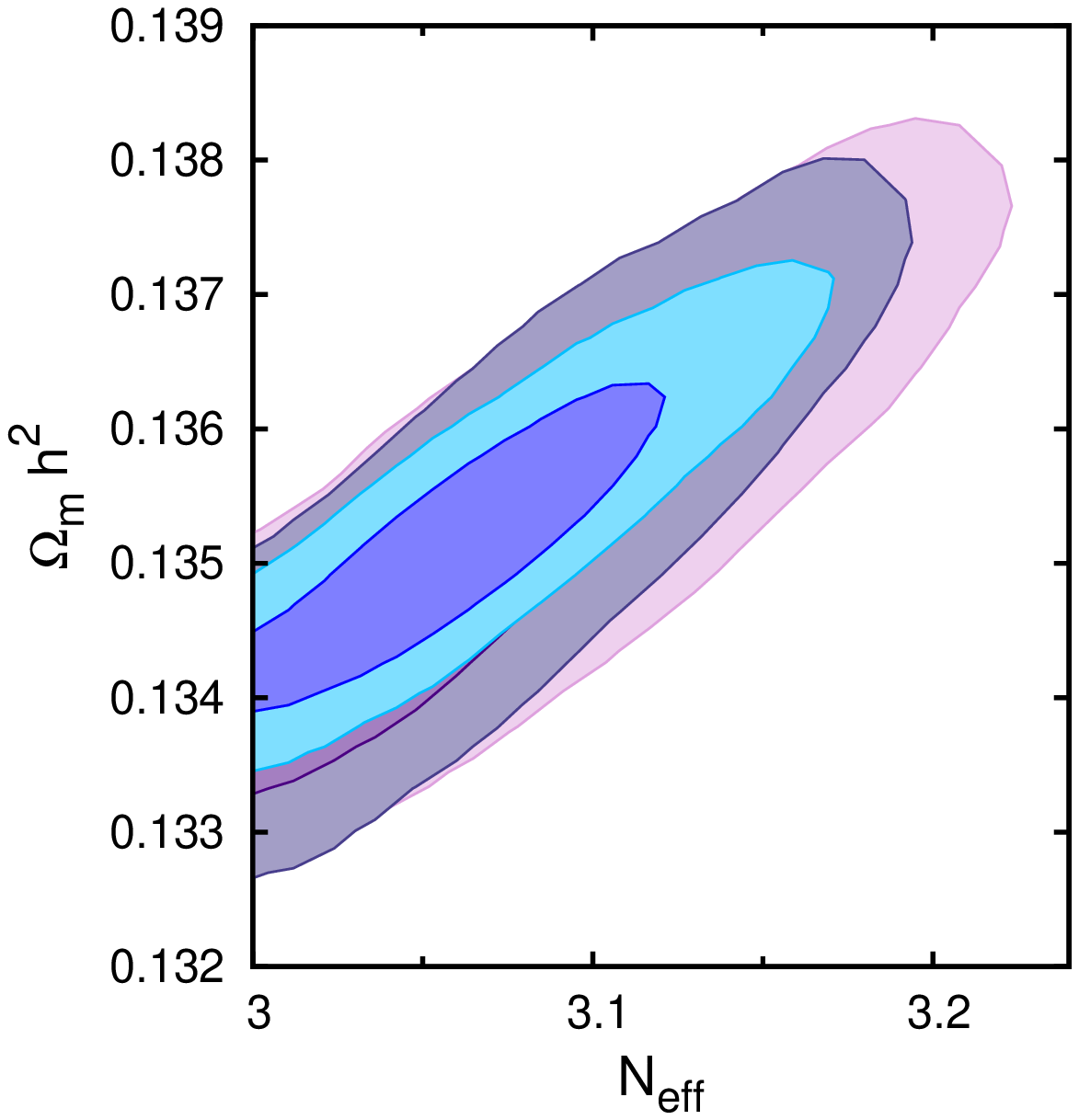}
\caption{Joint two dimensional marginalized constraints on the planes
  $(\sum m_\nu-\sigma_8)$ and $(N_{\eff}-\Omega_{\rm m} h^2)$ at
  $68\%$ and $95 \%$ CL from Euclid-Cl+Planck data. The confidence
  regions are for the $\Lambda$CDM+$m_\nu$+$N_{\eff}$ model discussed
  in~\ref{ssub:neff} (small blue contours) and the two extended model
  with nuisance parameters: all nuisance float (larger light violet
  contours) and fixed evolution parameters $\alpha$ and $\beta$ (dark
  violet contours).}
\label{fig:nuis}
\end{figure}

Finally, to assess the effect of an uncertain knowledge of cluster
masses on $\sum m_\nu$ and $N_{\eff}$ constraints, we treat the
$\Lambda$CDM$+m_\nu+N_{\eff}$ case with the four nuisance parameters
as fitting parameters, following the so-called self-calibration
method~\citep[e.g.][]{2004ApJ...613...41M, 2005PhRvD..72d3006L,
  sartoris:2011}. The results are listed in Table~\ref{tab:nuis}. We
start with the the over-conservative assumption of no priors on all
the four nuisance parameters
($\Lambda$CDM+$m_\nu$+$N_{\eff}$+$\text{nuis}$ model). The
uncertainties on scaling relation parameters compromise our ability to
recover the halo mass function from cluster data, thus reducing the
cosmological information achievable from cluster number counts. This
results in a larger error for the parameters that are primarily
constrained by cluster number counts, in particular for $\sigma_8$,
the normalization of the power spectrum. Looking at
Fig.~\ref{fig:nuis} (left panel), the constraints on $\sigma_8$ are
relaxed by a factor of $\sim 10$ compared to the
$\Lambda$CDM+$m_\nu$+$N_{\eff}$ model, and the parameter recovers a
large degeneration with $\sum m_\nu$ of the order of $\sim0.65$.  This
effect, along with the degradation of other parameters errors
(e.g. $\sigma(\Omega_{\rm m})$), entails a relaxation of the upper
limit for $\sum m_\nu$ by a factor larger than two, from $0.040\,
\text{eV}$ to $0.083\, \text{eV}$. With these loose constraints, in the
case of minimal normal hierarchy scenario, it would not be possible to
have a two $\sigma$ detection of neutrino mass.  Because the
constraints on neutrino mass from cluster number counts relay on the
evolution of the high-mass end of the mass function, $\sum m_\nu$ is
rather degenerate with $\alpha$ and $\beta$, the two nuisance
parameters which control the evolution of the systematic bias and
intrinsic scatter (see Eq.~\ref{eq:nuis}). To emphasize the role
played by the uncertain redshift evolution of the nuisance parameter
on the determination of $\sum m_\nu$ we show in Fig.~\ref{fig:nuis}
the contours for a model with $\alpha$ and $\beta$ kept fixed
($\Lambda$CDM+$m_\nu$+$N_{\eff}$+$\text{nuis}_{ev}$ model). In this
case the degradation of the total neutrino mass constraints with
respect the $\Lambda$CDM+$m_\nu$+$N_{\eff}$ model is only of $\sim
40\%$, from $0.040\, \text{eV}$ to $0.056\, \text{eV}$. In other
words, an accurate knowledge of the redshift evolution of the nuisance
parameter improves the $2\sigma$ upper limit of $\sum m_\nu$ by $\sim
33\%$ compared to the previous case with no prior on the nuisance
parameters.

Likewise, the forecast error on $N_{\eff}$ is influenced by the loss
of constraining power of the cluster number counts data, even if to a
lesser extent than the bounds on $\sum m_\nu$, since the constraints on $N_{\eff}$ are primarily contributed
by CMB measurements. The $2\sigma$ upper limit shifts from
$N_{\eff}<3.14$ to $N_{\eff}<3.16$ and $N_{\eff}<3.18$, in the model
with strong evolution prior and free nuisance parameters,
respectively. In these case the degradation is mainly due to the
larger error associated to $\Omega_{\rm m} h^2$, which is highly degenerate
with $N_{\eff}$ as explained in section~\ref{ssub:neff} and shown in
Fig.~\ref{fig:nuis} (right panel).  We remind that the results for the model
with no prior have to be regarded as an upper limit on the error
introduced by the uncertain knowledge of the scaling relation;
nevertheless, these results highlight the importance of having robust
calibration of the scaling relation, and in particular of their
evolution with redshift, to fully exploit the cosmological information
contained in the Euclid cluster catalog.

\section{Conclusions}\label{sec:conc}
In this paper, we presented forecasts on the capability of a future
photometric galaxy cluster survey such as Euclid, in combination with
Planck-like data, to provide constraints on neutrino
properties. Specifically, we rely on two observables: the cluster
number counts and their power spectrum. Our analysis is based on the
Markov Chain Monte Carlo methods rather than the Fisher Matrix
technique, which results in more reliable error bars. We start by
considering a reference $\Lambda$CDM model in agreement with the
recent results of WMAP 9-yr.  

In order to study possible degeneracies with $\sum m_\nu$, besides the
$\Lambda$CDM model with massive neutrino, we also consider models with
$N_{\eff}$ effective number of relativistic species, a constant dark
energy equation of state $w$ and curvature. Following the
self-calibration approach, along with the other cosmological
parameter, we decide to explore also the effect of leaving free the
nuisance parameters that describe the relation between cluster optical
richness and mass, its scatter and redshift evolution.

Our results can be summarized as follows:
\begin{itemize}

\item From the combination of Euclid number counts and clustering data
  we obtain a $2\sigma$ upper limits for the total neutrino mass of
  $\sum m_\nu<0.35 \,\text{eV}$, comparable with present constraints
  from the combination of CMB and LSS
  probes~\citep[e.g.][]{2012arXiv1212.5226H,2012arXiv1211.3741Z,2012AstL...38..347B}.
  When Planck data are added to the Euclid-Cl ones the error on $\sum
  m_\nu$ is reduced by a factor larger than $10$ to $\sum
  m_\nu<0.031\,\text{eV}$. With this accuracy the total neutrino mass
  could be detected at $2\sigma$ level even in the minimal normal
  hierarchy scenario. The large improvement is due to the different
  degeneracies present between Euclid and Planck that are broken once
  the two experiments are combined.

\item Because the effective number of neutrino spices is degenerate
  with the sum of neutrino masses, varying $N_{\eff}$ entails a
  relaxation of $2\sigma$ error bars on $\sum m_\nu$ by $\sim 30\%$ in
  the Planck+Euclid-Cl case. Still, the $2\sigma$ error is lower than
  the minimum neutrino mass admitted by neutrino oscillation
  experiments.  The Euclid-Cl dataset is unable to constraints
  $N_{\eff}$ by itself, but improves the $2\sigma$ upper limits on
  $N_{\eff}$ from $3.36$ using Planck-only, to $3.14$ in the
  Planck+Euclid-Cl case. The improvement is mainly due to the tighter
  constraints on $\Omega_{\rm m} h^2$ provided by the Euclid-Cl
  datasets.

\item In models with varying $w$ or $\Omega_{\rm k}$ the $2\sigma$
  error on $\sum m_\nu$ is relaxed only by $\sim 10\%$.
  In both cases the high accuracy with which $w$ or
  $\Omega_{\rm k}$ are constrained by the Planck+Euclid-Cl data
  prevents the error on $\sum m_\nu$ from being largely degraded.  As
  for $N_{\eff}$, the parameter shows a correlation of the order of
  $\sim 0.5$ with both $w$ and $\Omega_{\rm k}$, which shifts the
  $2\sigma$ upper limit for $N_{\eff}$ to $3.16$ and $3.17$,
  respectively.

\item When nuisance parameters are considered in a conservative way
  (no prior) our ability to recover the halo mass function from
  cluster data is compromised. The degradation of cosmological
  information results in a $\sim 2$ times larger $2\sigma$ error for
  neutrino masses ($\sum m_\nu<0.083\,\text{eV}$) and a degradation of
  $\sim 30\%$ of the $2\sigma$ error on the effective number of
  neutrinos ($N_{\eff}<3.18$).  In this case the accuracy would not be
  sufficient for detecting the total neutrino mass with good
  significance in the minimal normal hierarchy scenario. Whereas,
  assuming a perfect knowledge of the redshift evolution of the
  nuisance parameters we partially recover the informations contained
  in cluster number counts data. In this case the $2\sigma$ upper
  limit for $\sum m_\nu$ is degraded only by $40\%$ to $\sum
  m_\nu<0.056\,\text{eV}$, while the $2\sigma$ error on the effective
  number of neutrinos degrades by $\sim 15\%$ to $N_{\eff}<3.16$.
 \end{itemize}
 It is worth reminding that in our analysis we did not include the
 effect of redshift space distortions in the distribution of galaxy
 clusters induced by peculiar velocities. This effect should be in
 principle included when forecasting the cosmological constraining
 power of future cluster surveys; indeed, as demonstrated
 by~\cite{sartoris:2011}, the inclusion of redshift space distortions
 carries significant cosmological information through the growth rate
 of density perturbations.

 As a concluding remark, we emphasize once again the importance to
 provide an accurate calibration of the scaling relation between the
 observable quantity on which cluster selection is based, optical
 richness in this case, and cluster mass. Thanks to the exquisite
 imaging quality expected for the Euclid survey, weak lensing masses
 for individual objects will be available for a significant fraction
 of the clusters identified from the photometric selection. At the
 same time, staking analysis will provide an accurate calibration of
 the relation between weak lensing masses and richness. For this
 reason, the Euclid cluster survey will represent a powerful
 complement to galaxy clustering and cosmic shear analyses to
 constrain cosmology through the growth of perturbations.

\textbf{Note added}: After the submission of this work cosmological results
from the first 15.5 months of Planck operations has been published~\cite{2013arXiv1303.5076P}.
The fiducial values adopted in this work are found to be consistent within $2\sigma$ 
with the mean values obtained by Planck Collaboration.
For this data release the authors did not use polarization spectra, so we can not make
a direct comparison of our forecast with the actual constraints from Planck.
However, for a $\Lambda$CDM+$m_\nu$ model, using Planck temperature power spectrum in combination with a WMAP-9yr
polarization low-multipole likelihood the authors obtained $\sum m_\nu < 0.933$ ($95\%$CL),
compatible with our expected error.

\section*{Acknowledgments}
We thank Lauro Moscardini and Jochen Weller for useful
comments, and an anonymous referee for constructive criticisms that
helped improving the presentation of the results.
We acknowledge financial support from the agreement ASI/INAF I/023/12/0.
 This work has been supported by the PRIN-INAF09 project
``Towards an Italian Network for Computational Cosmology'', by the
PRIN-MIUR09 ``Tracing the growth of structures in the Universe'', and
by the PD51 INFN grant. MV is supported by the ERC Starting Grant
CosmoIGM.  JX is supported by the National Youth Thousand Talents
Program and the grants No. Y25155E0U1 and No. Y3291740S3.

\bibliographystyle{utcaps}
\bibliography{Bibliography}

\end{document}